\documentclass[journal]{IEEEtran}  

\usepackage[]{graphicx,float,latexsym,amssymb,amsfonts,amsmath,amstext,times}
\usepackage{cite}
\newcommand{\ignore}[1]{}  
\usepackage{amsmath}
\usepackage[hyphens]{url}
\usepackage[]{graphicx}    
\usepackage[T1]{fontenc}
\usepackage{array}
\newcolumntype{P}[1]{>{\centering\arraybackslash}p{#1}}
\usepackage{multirow}
\usepackage{multicol}
\usepackage{mathtools}
\usepackage{graphicx}
\usepackage{balance}

\begin{document}
\title{Micro-UAV Detection and Classification from RF Fingerprints Using Machine Learning Techniques}

\author{%
Martins Ezuma, Fatih Erden, Chethan Kumar Anjinappa, Ozgur Ozdemir, and Ismail Guvenc\\ 
Department of ECE\\
North Carolina State University\\
Raleigh, NC 27606\\
\{mcezuma,ferden,canjina,oozdemi,iguvenc\}@ncsu.edu
}

\maketitle

\thispagestyle{plain}
\pagestyle{plain}

\maketitle

\thispagestyle{plain}
\pagestyle{plain}


\begin{abstract}
 
This paper focuses on the detection and classification of micro-unmanned aerial vehicles (UAVs) using radio frequency (RF) fingerprints of the signals transmitted from the controller to the micro-UAV. 
In the detection phase, raw signals are split into frames and transformed into the wavelet domain to remove the bias in the signals and reduce the size of data to be processed. A naive Bayes approach, which is based on Markov models generated separately for UAV and non-UAV classes, is used to check for the presence of a UAV in each frame. In the classification phase, unlike the traditional approaches that rely solely on time-domain signals and corresponding features, the proposed technique uses the energy transient signal. This approach is more robust to noise and can cope with different modulation techniques. First, the normalized energy trajectory is generated from the energy-time-frequency distribution of the raw control signal. Next, the start and end points of the energy transient are detected by searching for the most abrupt changes in the mean of the energy trajectory. Then, a set of statistical features is extracted from the energy transient. Significant features are selected by performing neighborhood component analysis (NCA) to keep the computational cost of the algorithm low. Finally, selected features are fed to several machine learning algorithms for classification. The algorithms are evaluated experimentally using a database containing 100 RF signals from each of 14 different UAV controllers. The signals are recorded wirelessly using a high-frequency oscilloscope. The data set is randomly partitioned into training and test sets for validation with the ratio 4:1. Ten Monte Carlo simulations are run and results are averaged to assess the performance of the methods. All the micro-UAVs are detected correctly and an average accuracy of 96.3\% is achieved using the k-nearest neighbor (kNN) classification. Proposed methods are also tested for different signal-to-noise ratio (SNR) levels and results are reported.
\end{abstract}



\section{INTRODUCTION}


In recent years, non-military micro-unmanned aerial vehicles (micro-UAVs) have proliferated conspicuously. Besides the recreational use by hobbyists, there is a growing interest in the use of micro-UAVs for commercial applications. One of the major areas of use is in precision agriculture, 
where micro-UAVs make it easy to map and survey farmlands for crop variability and phenology, crop dusting/spraying for weed and pest control, irrigation management, and livestock monitoring~\cite{FAOUnitedNations}. Other commercial applications of micro-UAVs include infrastructure health monitoring, package delivery, media \& entertainment, 
and ad hoc access point Internet connectivity~\cite{kangautonomous,ackerman2018medical,gatteschi2015new,shi2018multiple}. 
Due to the potential benefits of micro-UAVs, there is a collaborative plan by the Federal Aviation Administration (FAA) and National Aeronautics and Space Administration (NASA) to integrate commercial micro-UAVs into the national airspace (NAS)~\cite{nasautm}.
  
Even though so many beneficial civilian applications of micro-UAVs abound, there is an associated risk to the public safety. In recent times, there have been reports of micro-UAVs violating public privacy and the security of sensitive facilities such as nuclear power plants and airports~\cite{solodov2018analyzing}. In 2018, a drone was intentionally crashed into a nuclear power plant in France ~\cite{FAA_news2}. According to the FAA, reports of safety-incidents involving drones now average about 250 a month~\cite{FAA_news}. Some of these events involve micro-UAVs crashing into commercial airplanes, military helicopters, the White House, and outdoor public events. Apparently, most of these events occur when drone pilots intentionally violate no-fly-zone restrictions. 
In addition, micro-UAVs have been exploited by terror groups for the placement of improvised explosive devices (IED) and chemical, biological, radiological, nuclear, and explosives (CBRNE)~\cite{hutter2016possibilities}. Recently, two armed commercial drones carrying powerful explosives detonated close to the Venezuelan president during an outdoor event~\cite{Venezuelan_drone_incident}. 
Therefore, there is an urgent need to secure the national airspace against such unconventional threats. This can be achieved by accurately detecting and identifying non-compliant micro-UAVs.
   
Several techniques have been proposed for micro-UAV detection and classification so far. Conventional radar-based techniques, which are widely deployed for detecting and identifying aircrafts, mostly fail to detect micro-UAVs~\cite{White_House}. 
Alternative techniques like sound and video-based detection are only suitable for short-range scenarios due to ambient noise~\cite{bisio2018unauthorized}. Some of these challenges can be addressed by radio frequency (RF) fingerprints-based techniques. However, the current trend on RF fingerprint classification of micro-UAVs focuses mostly on time-domain techniques which are not very effective. This is because time domain techniques are based on the assumption that there is an abrupt change at the start point of the signal. However, this assumption is not always true when the transition between the transient and noise is more gradual  \cite{Hall2013}. Consequently, time domain techniques may delay the detection of the transient of the signal. In worst case scenario, this may increase the probability of missed target detection at low signal-to-noise-ratio (SNR).

This paper is motivated by the need to address the aforementioned challenges. Due to the problems associated with the use of the time-domain transient analysis, a new approach for the micro-UAV classification is proposed in this paper. In this approach, the time-domain signal is first transformed into the energy-time-frequency domain and the energy trajectory is computed. Then, a set of statistical features is extracted from the energy transient instead of the time-domain transient. The dimensionality of the feature set is reduced using neighborhood component analysis (NCA) and the significant features are classified using several machine learning algorithms. It is shown that the  discriminating features can still be extracted even when the time-domain signal waveform is distorted by noise. Moreover, a micro-UAV detection method is described in this paper. RF signals are transformed into the wavelet domain to remove the bias and reduce the size of the data. Then, the naive Bayes approach is used based on the Markov models to differentiate between the noise and micro-UAV signals. 

The paper is organized as follows. Section~\ref{two} gives an overview of the state-of-the-art techniques for micro-UAV detection and classification; Section~\ref{two_1} provides a description of the process model for the approach;  Section~\ref{three} and Section~\ref{four_1} describes the proposed detection and classification techniques respectively; Section~\ref{four} describes the experimental setup and presents the results; and Section~\ref{five} provides the concluding remarks.

\section{RELATED WORK} \label{two}
Existing techniques for micro-UAV detection and classification can be categorized under four headings, namely, radar-based, vision-based, sound-based, and RF fingerprinting. 
\subsection{Radar-based Techniques}
Micro-UAV detection using radars have been widely studied. Radars transmit electromagnetic signals which interact with the target, in particular, the micro-UAVs. This interaction causes a shift in the carrier frequency of the received signal due to the Doppler effect. In addition to the main Doppler shift, if the target has vibrating or rotating structures, e.g., propellers, vibrating platforms, and engines, these micro-motions will induce time-varying frequency modulation on the received signal~\cite{chen2011micro}. These additional frequency modulations, called the micro-Doppler effect, generate side-bands (or spectral lines) around the main Doppler frequency shift. Analysis of the micro-Doppler signature may review some of the dynamic characteristic of the target that can be used for target detection and identification~\cite{zhang2016micro,thayaparan2004micro}. 
  
In~\cite{MartinsDrone}, the micro-Doppler signature of a quad-copter is compared with that of a walking human. The study concludes that the unique micro-Doppler signature of the micro-UAVs are useful for the design of an automatic target recognition (ATR) system. These unique features can be used to distinguish micro-UAVs from fixed-wing airplanes, helicopter, and birds. 
In~\cite{molchanov2014classification}, micro-UAVs and small birds are classified based on the eigenpairs extracted from the decomposition of their micro-Doppler signatures. Larger birds can be readily recognized and discriminated from small-UAVs because of the frequency modulation induced by their flapping wings~\cite{torvik2014amplitude,torvik2014x}. These flapping-induced micro-Doppler frequencies appear at a much lower  frequency-band as compared to the micro-Doppler frequencies induced by the rotating propellers of the micro-UAVs. Sparsity-based techniques can also be used to extract features from radar micro-Doppler signatures. In~\cite{vishwakarma2016classification,li2017sparsity}, orthogonal matching pursuit (OMP), a sparse-coding based dictionary learning algorithm, is used to extract features from the radar micro-Doppler signatures for automatic target recognition.

Although radar-based detection has been one of the mainstream approaches to the problem, their performance, i.e., maximum detectable range, is highly limited if the target has a low radar cross-section~\cite{guvencc2017detection}. This also explains why stealth aircrafts, designed to avoid radar detection, must have very low radar cross section (RCS). A conventional aircraft (without stealth coating) has an average RCS value of 426.58 m\textsuperscript{2} (26.3 dBsm) at lateral incidence to a millimeter wave radar signal~\cite{dos2014analysis}. Therefore, in order to evade detection, military aircrafts are coated with radiation-absorbent materials (RAM) to reduce the RCS. On the other hand, experimental measurements at millimeter wave frequencies show that many commercially available micro-UAVs have an average RCS value of about 0.02 m\textsuperscript{2} (-16.98 dBsm)~\cite{li2017investigation,dedrone,nakamura2017characteristics}. This very low RCS value is due to the shape and design material of these micro-UAVs. Therefore, many micro-UAVs are naturally stealth to conventional radars~\cite{guvencc2017detection}. This explains the failure of the United States White House's surveillance radar to detect a micro-UAV flying across the fence and crash-landing into the lawn~\cite{White_House}. This challenge with radar-based detection has motivated researchers to investigate other detection techniques for micro-UAVs.

\subsection{Vision-based Techniques}
In ~\cite{gokcce2015vision}, a computer vision-based technique is described for micro-UAVs detection. This approach uses high resolution cameras to capture micro-UAVs in different background environments. Several features such as Haar-like, histogram of gradients (HOG), and local binary patterns (LBP) are extracted from the images. These features are fed into cascades of boosted classifiers for target detection. The cascaded boosted classifiers perform detection at multi-stage sequences with increasing complexity. In this system, only test sets that pass the previous stage are allowed into the next stage. 

Deep learning networks have also been explored for the micro-UAV detection problem. Usually, deep learning techniques do not rely on the human crafted features for target detection. They autonomously learn the optimal features from the captured micro-UAV images. In~\cite{aker2017using,pengusing}, convolution neural networks (CNN) are investigated for micro-UAV detection. These deep learning based techniques show fairly good performance. However, training CNN networks requires huge amount of data making real-time application computationally expensive. 

In~\cite{unlu2018using,unlu2018generic}, the authors described a computer-vision approach based on a generic Fourier descriptor (GFD). The technique uses speeded-up robust features (SURF) for keypoint detection on grayscale images of micro-UAVs. The keypoints of interest are shape-descriptors of the micro-UAVs. In comparison to CNN, this method provides much faster detection for micro-UAVs. However, all vision-based techniques suffer from one common problem. The performance of the camera sensors depends on the ambient condition such as lighting. In addition, vision-based detection of micro-UAVs may not perform well if the surveyed area is large.

 \subsection{Sound-based Techniques}
  \begin{figure*}[t!]
\center{\includegraphics[scale=0.65]{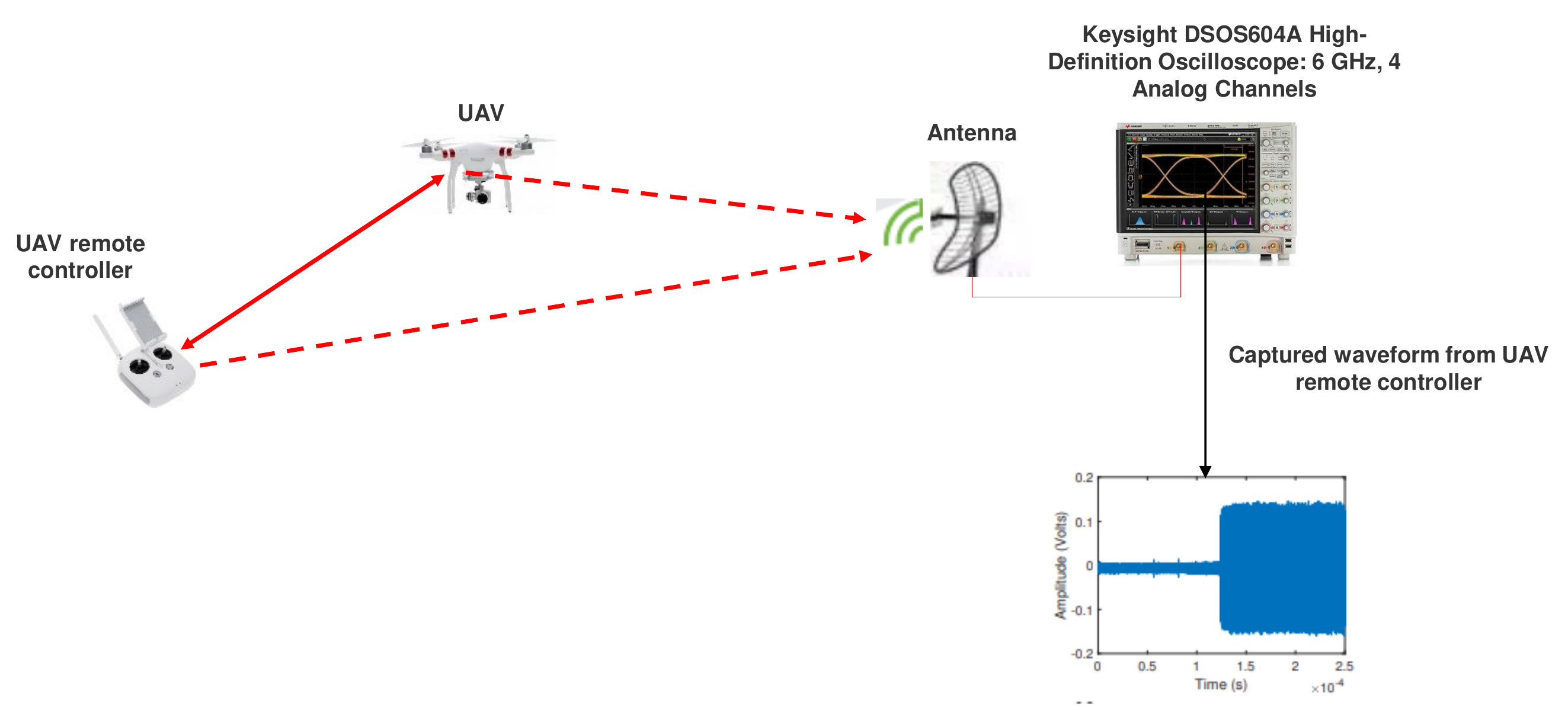}}
\caption{System setup for RF-based UAV detection.}
\label{Fig:RFsystemsetup}
\end{figure*}

 Acoustic or sound-based techniques use arrays of micro-phones to extract the unique acoustic signature of micro-UAVs. Typically, micro-UAVs produce hissing or buzz-like sounds in frequencies ranging from 400 Hz to 8 kHz~\cite{Student_thesis-Lund}. This unique acoustic signature is due to the brushless DC motor of micro-UAVs. Using different audio analysis techniques, micro-UAVs can be separated from the background noise. In~\cite{chang2018surveillance}, micro-UAV localization and tracking using an acoustic array is described. The micro-UAVs are localized based on estimation of the time difference of arrival (TDOA) of the received audio signals. In order to accurately compute TDOA, the authors proposed an algorithm
based on the Gauss priori probability density function
(GPDF). 

In~\cite{bernardini2017drone}, time and frequency domain acoustic features are extracted from micro-UAV audio recordings. These features are used to train a multi-class support vector machine (SVM) for micro-UAV identification. In~\cite{jeon2017empirical}, the authors investigate the effectiveness of Gaussian mixture model (GMM) and deep learning algorithms for drone sound detection. The problem, modeled as a binary classification problem, is based on the detection of sound events. The study concludes that a long short-term memory (LSTM) recurrent
neural network (RNN) shows the best micro-UAV sound detection performance. In ~\cite{busset2015detection}, micro-UAV detection using hybrid advanced acoustic cameras is described. The system comprises 120 elements microphone array and a video camera. The microphones are spherically arranged. Thus, allowing the system to simultaneously detect multiple micro-UAVs in 2D (angular position) or 3D dimensions. The angular direction of a micro-UAV is estimated using the phase of the acquired audio signals from the micro-UAVs. 

In~\cite{liu2017drone}, a similar hybrid audio-assisted detection system for micro-UAVs is described. The system consists of thirty high-definition cameras and three microphones. In order to perform micro-UAV detection, HOG features are extracted from image data and mel-frequency cepstral coefficients (MFCCs) features from the audio data. An SVM is trained to perform detection on the test data set. The fundamental challenge of the audio-based systems is the practical range of the commercial microphones. Most of them have a range of 25-30 ft~\cite{bisio2018unauthorized} and are highly sensitive to environmental noise.

 \subsection{RF fingerprinting}

RF fingerprint-based detection relies on the characteristics of the RF signals of the micro-UAV controllers. Experimental investigations show that the micro-UAV controllers have a unique RF signature due to the circuitry design and modulation techniques employed. Therefore, RF fingerprint analysis can help detect and classify micro-UAVs. Unlike the radar-based techniques, the RF sensing device/receiver is a passive listener and does not transmit any signal. This makes RF fingerprint-based detection energy efficient. In addition, the challenge of detecting micro-UAVs (with extremely low RCSs) is solved since all that is required is to intercept the transmit signal from the micro-UAV controller. The range problem associated with the vision and acoustic-based techniques can be solved by using high-gain receiver antennas together with a highly sensitive receiver system to listen for the micro-UAV controller signals. The issue of environmental noise can be suppressed by employing several de-noising techniques, e.g., wavelet decomposition and band pass filtering. These advantages make RF fingerprint detection techniques a promising solution.

In~\cite{Caidan}, GMMs are to detect the transient start points of signals transmitted by a micro-UAV controller. This time-domain technique uses the expectation maximization (EM) algorithm for estimating the detection threshold. However, EM algorithm is not guaranteed to converge to a global optimal. Furthermore, in order to justify the use of GMM to model the RF signal sampled with a high frequency, many Gaussian components are needed. This will definitely increase the computational cost of the detection algorithm. In~\cite{Caidan_hash}, RF hash fingerprints are used to detect the micro-UAVs. The RF hash fingerprints are generated by 
calculating the distance between peak locations in the envelope of the time-domain RF signal. The extracted fingerprints are used to train a distance-based support vector data description (SVDD)
algorithm for target detection. However, since most micro-UAV controller signals have similar time-domain waveforms with random spikes, time domain peak classification is not very effective.



\section{UAV DETECTION/CLASSIFICATION SCENARIO AND ASSUMPTIONS}\label{two_1}

\begin{figure}[h!]
\center{\includegraphics[scale=0.6]{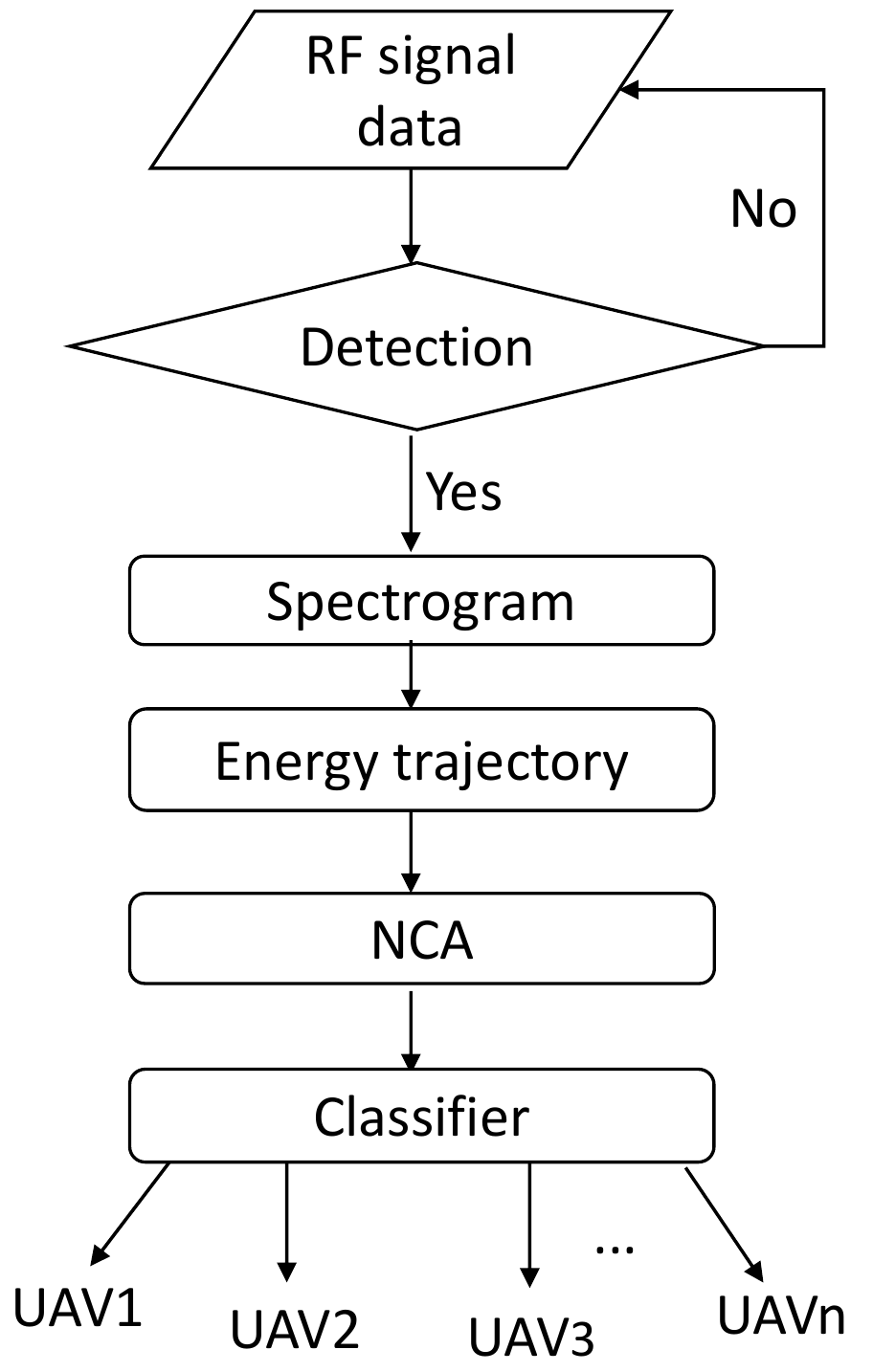}}
\caption{Flowchart of the classification algorithm.}
\label{Fig:flowchart}
\end{figure}

Fig.\ref{Fig:RFsystemsetup} shows the system setup for RF-based UAV detection and identification. The detection system is an RF sensing device which can capture signals from both the UAV and its controller. In addition, the RF receiver also captures other signals in the environment which co-exist with the UAV transmissions in the same frequency band. Our overall goal in this paper is to develop an algorithm which is capable of detecting a micro-UAV and, if present, identifying the type of it based on the extracted RF fingerprint of the UAV controller signal. Thus, the process is divided into two major tasks, namely, the detection and classification. The detection stage makes a decision of whether the captured RF signal belongs to a micro-UAV or the noise signal. If a micro-UAV is detected, then the classification stage is invoked to make a decision regarding the type of the micro-UAV. Fig.~\ref{Fig:flowchart} is a flowchart that provides a graphical description of the overall process model that involves UAV detection and classification.

The starting point for target classification is the target detection. This is achieved by continuously sensing the 2.4 GHz channel for the presence of transmissions by a non cooperative UAV controller. Experimentally, it was observed that it is much effective to detect the signal from the UAV controller as against the transmission from the UAV itself because the former has higher energy than the latter. For this work we focus exclusively on the detection/classification of the RF signal from the controller; RF signal detection/classification from the UAV is left as a future work.
  
Contrary to the classical energy detector system, which sets a single fixed threshold for target detection, the proposed system is based on state transition probabilities. This approach reduces the probability of false alarm due to random burst in the background noise. Moreover, this approach is motivated by the fact that actual RF signal waveforms from UAV controllers can be considered as a time-varying spectral vector sequences with multiple transients.
  
Fig.~\ref{fig:UAV data waveforms} shows typical RF signal waveforms captured from six different micro-UAV controllers. These waveforms look distinct with well defined transitions. Therefore, Markov based models can be effectively used to detect the UAV controller signal from the background noise. In this work, the background noise is modelled as an additive white Gaussian noise. According to \cite{pace2009detecting}, the maximum intercept range of the RF sensing system can be modelled as:
 \begin{equation}
    R_{\rm Imax} = \frac{\lambda}{4\pi}\sqrt{\frac{P_{\rm t} G_{\rm t}G_{\rm I}}{L\delta_{\rm I}}}~,
\end{equation}
where $\lambda$, $P_{t}$ and $G_{t}$ represents the transmit wavelength, transmit power and antenna gain of the UAV controller respectively, $G_{I}$ is the antenna gain of the RF receiver (intercept) system, $L$ is the combined losses between the controller and receiver (transmission and system losses) and $\delta_{I}$ is the sensitivity of the receiver. Moreover, the RF receiver sensitivity, $\delta_{\rm I}$, is defined as:
\begin{equation}
    \delta_{\rm I} = kT_{o}FB\rho_{i}~,
\end{equation}
where $k=1.38\times 10^{-23}$ Joules/k is Boltzmann constant, $T_{o}=290$ k is the standard noise temperature, $F$ is the noise factor of the receiver, $B$ is the bandwidth of the receiver and $\rho_{i}$ is the SNR at the input of the receiver. In practice, the maximum range can be increased by using directional antenna for far field detection.

The next task after signal detection is the classification process. This is achieved by using machine learning algorithms. First, the energy trajectory features of the captured RF signal waveform is extracted from the spectrogram representation of the signal. Thereafter, feature selection is then performed using Neighborhood component analysis (NCA). The selected features are used for training and testing of the machine learning models used for the classification. 
 
In the following, first, Section~\ref{three} will provide a detailed description of the proposed RF-based UAV detection approach, while Section~\ref{four_1} will focus on the UAV classification problem.

\section{UAV DETECTION USING RF SIGNALS}\label{three}

\begin{figure*}[t!]
    \center{
    \includegraphics[scale=0.35]{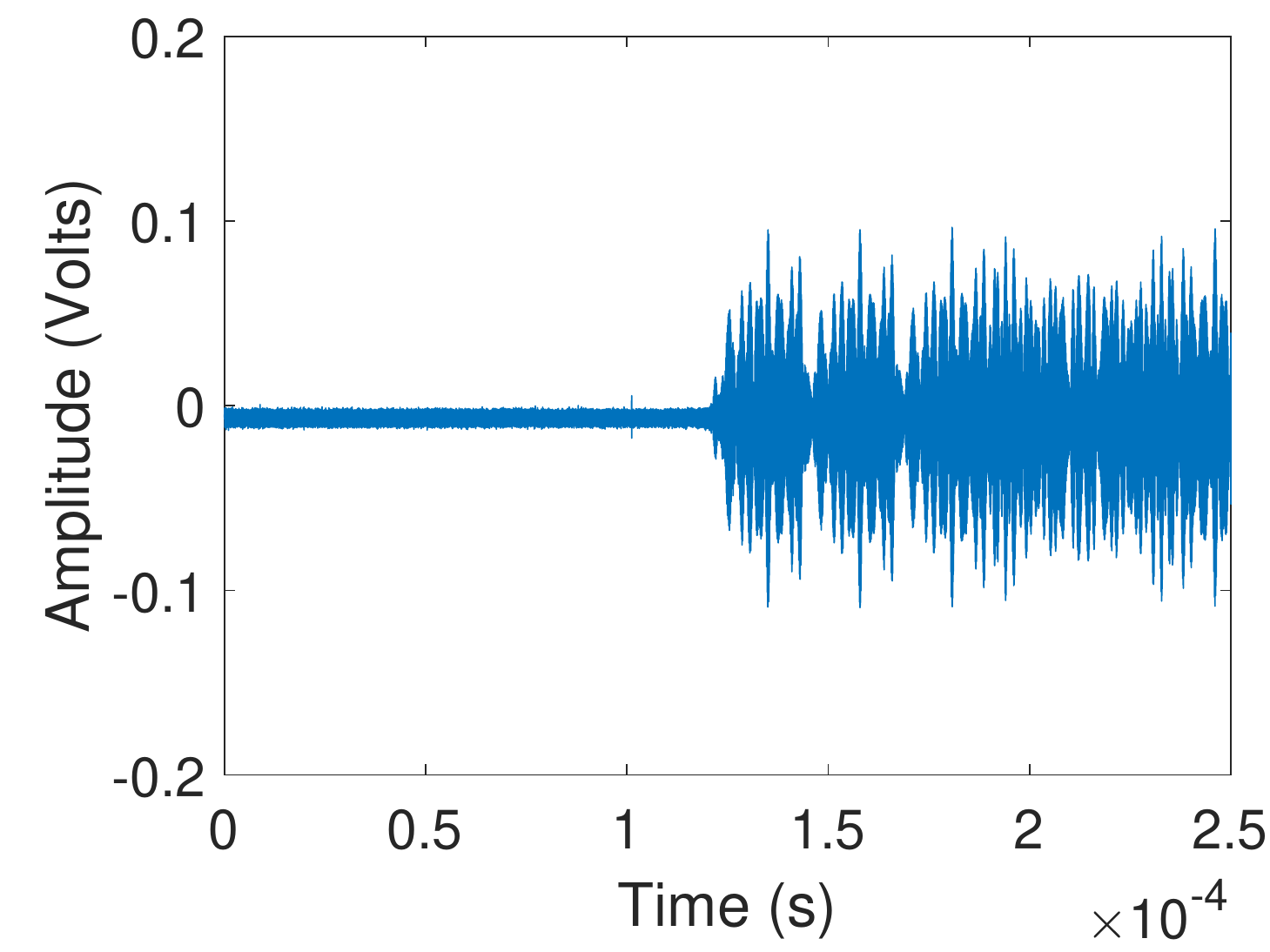} 
    \includegraphics[scale=0.35]{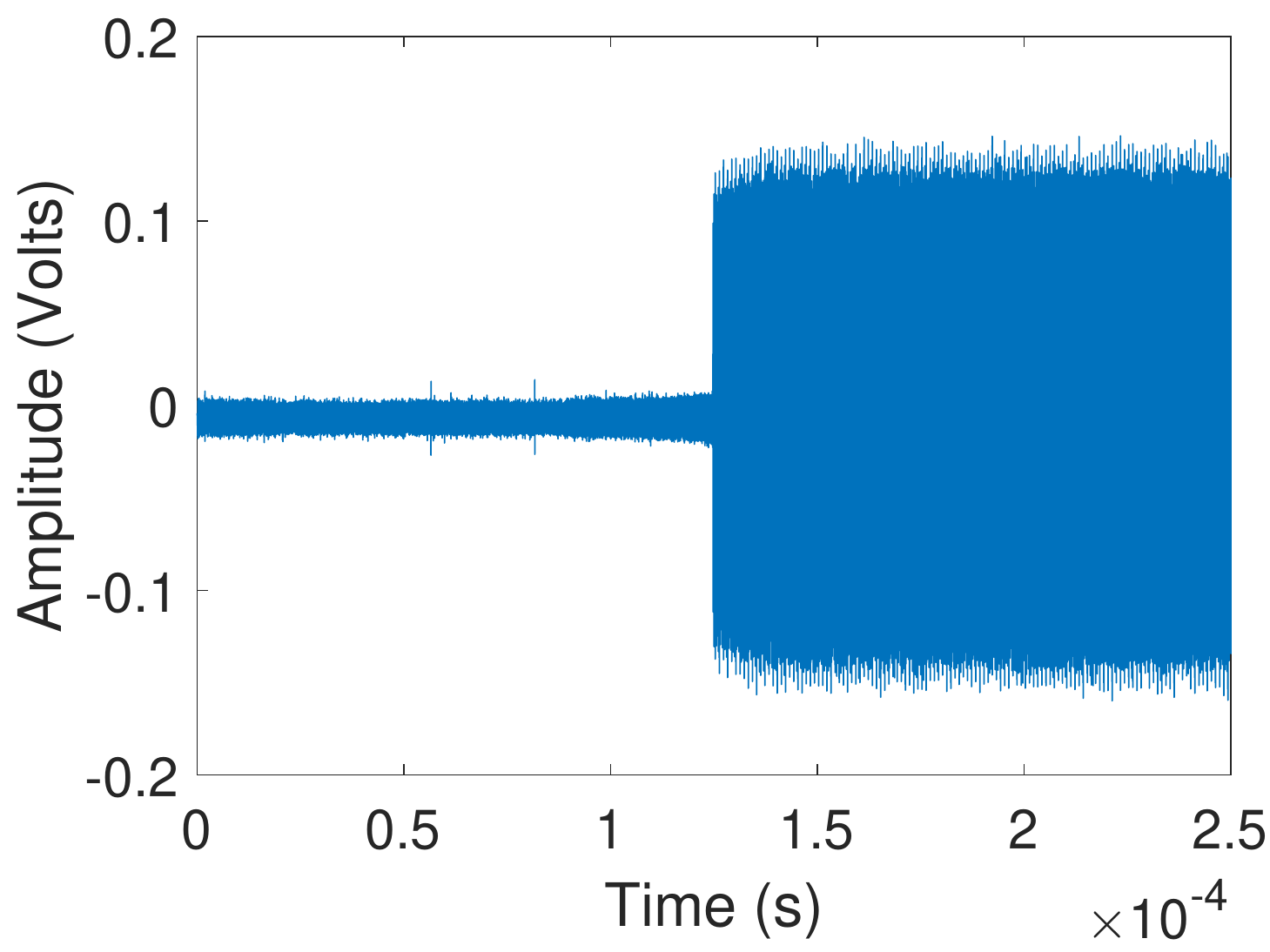}
    \includegraphics[scale=0.35]{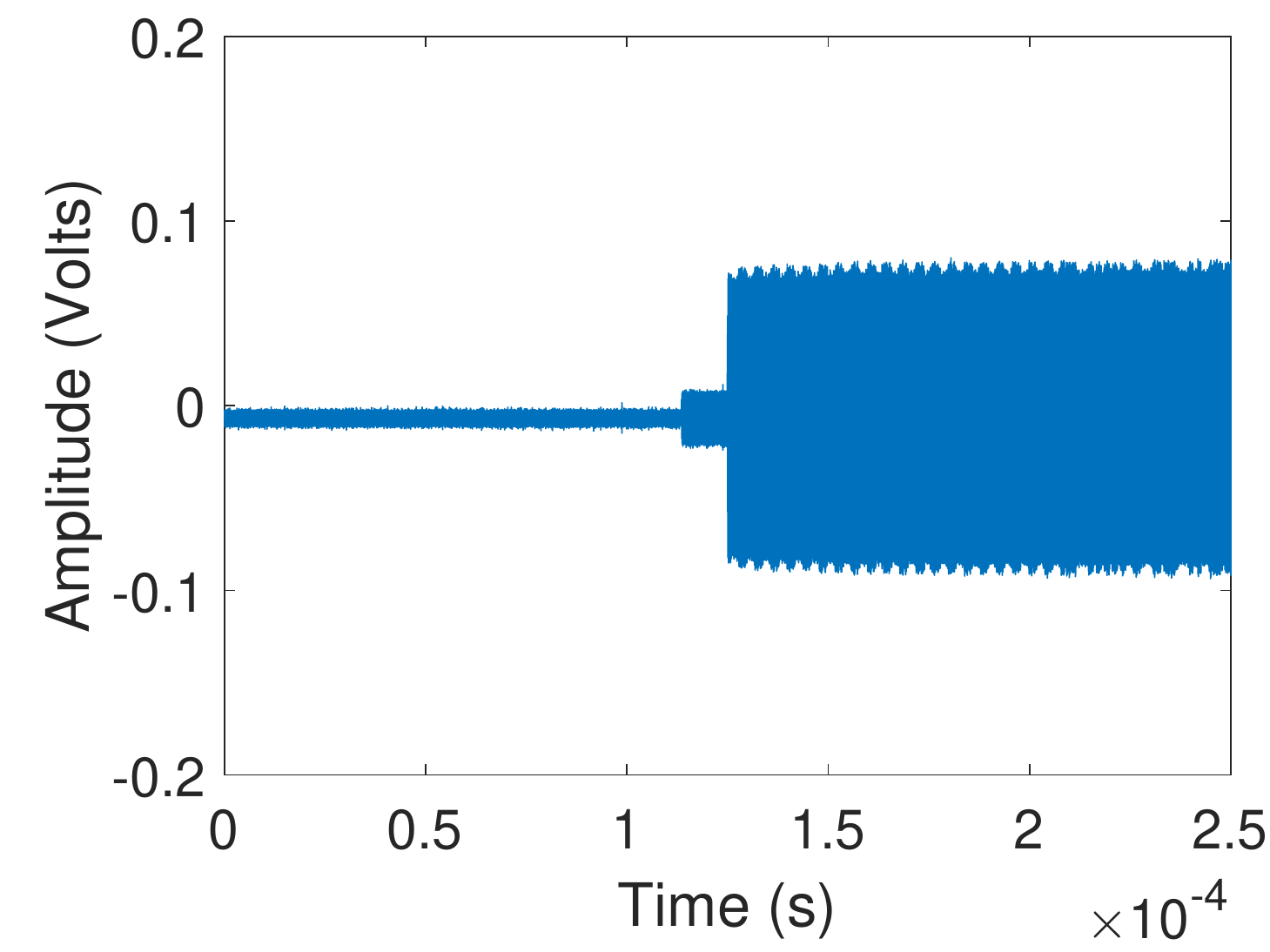}
    \includegraphics[scale=0.35]{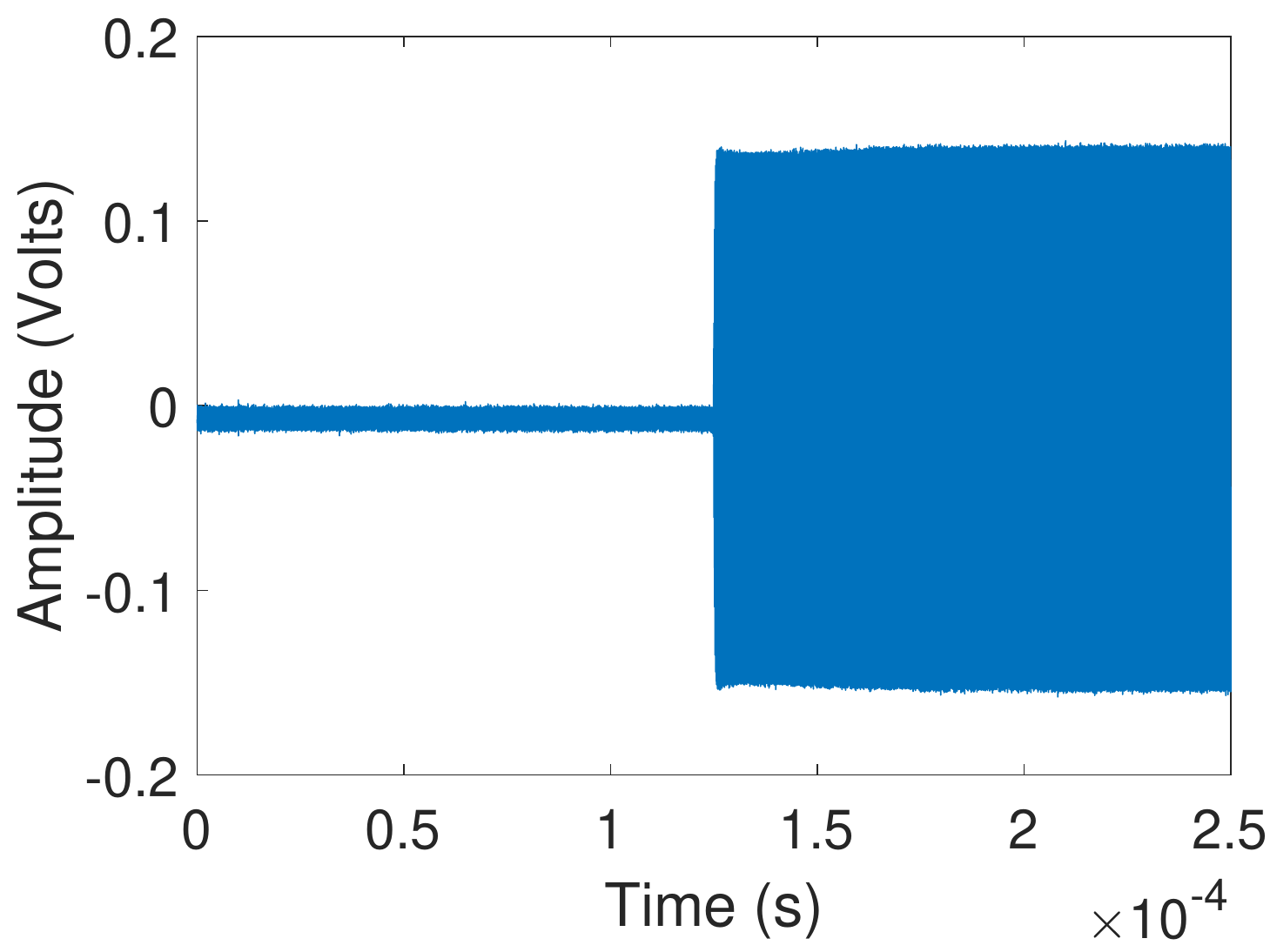} 
    \includegraphics[scale=0.35]{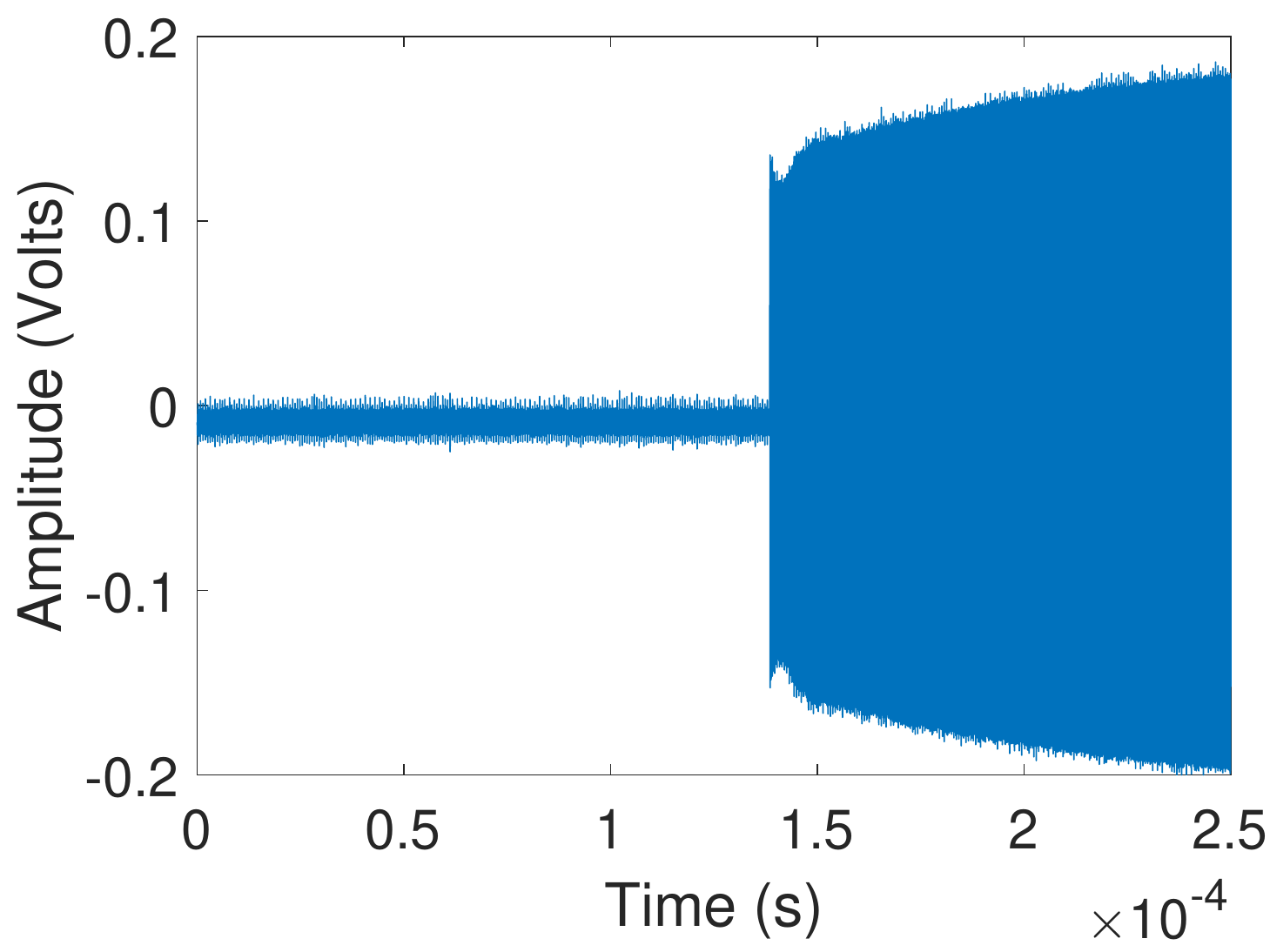}
    \includegraphics[scale=0.35]{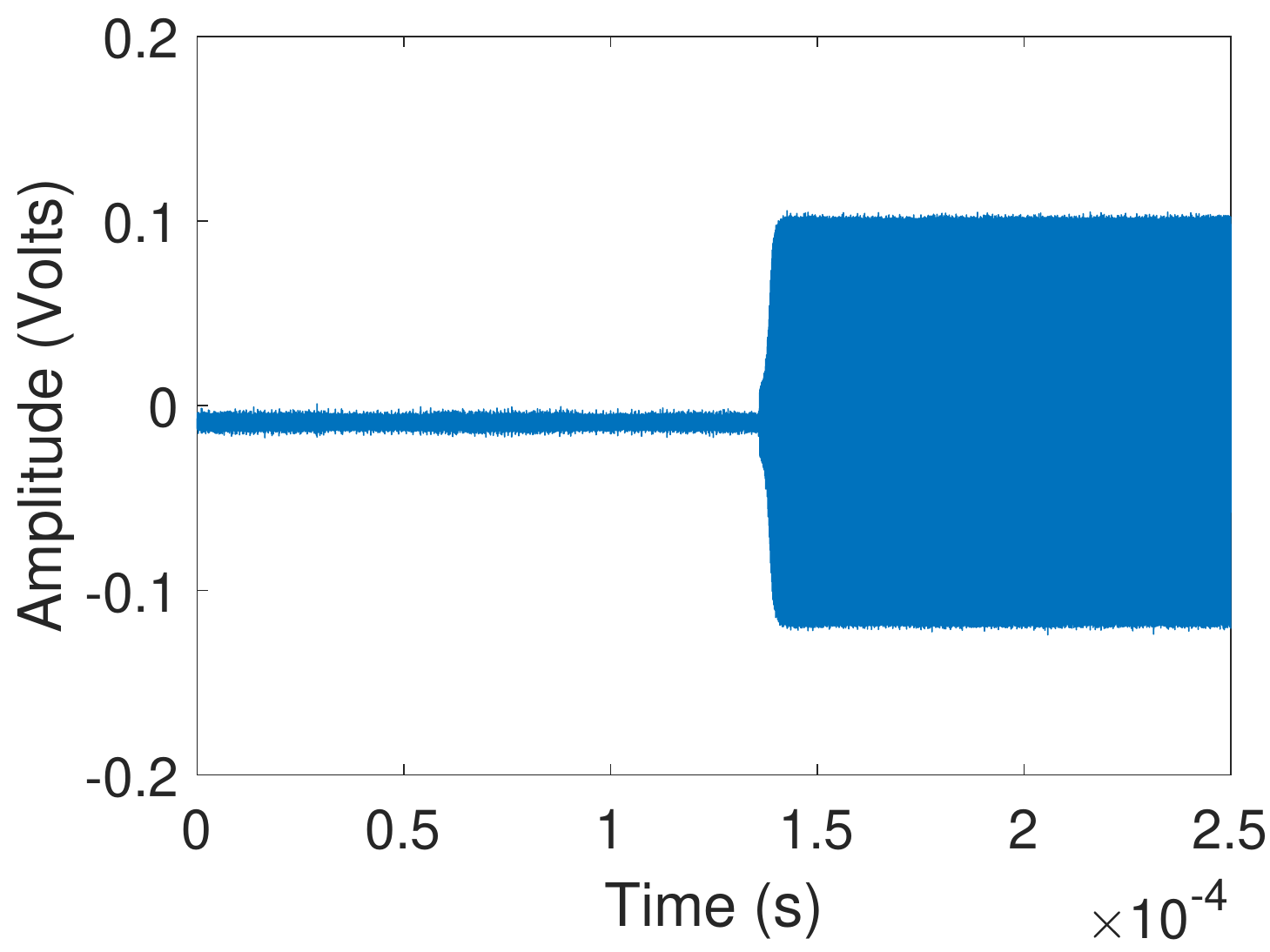}}
    \caption{{Examples of micro-UAV control signals for six different controllers: (a) DJI Matrice 100, (b) DJI Phantom 3, (c) Hobby King T6A V2, (d) DX6e Spektrum, (e) JR X9303, and (f) Jeti Duplex DC-16 (from top left to bottom right)}.}
    \label{fig:UAV data waveforms}
\end{figure*}

First step in Fig.~\ref{Fig:flowchart} is capturing RF signal data and the detection of the presence of a UAV signal in the data. 
In our experiments, the RF signals to be detected are captured from different micro-UAV controllers using a high-frequency oscilloscope. The details of the experimental setup and data collection are given in Section~\ref{four}. Each RF signal is recorded such that it is a vector of the same size. 

Fig.~\ref{fig:UAV data waveforms} shows the typical RF signals received from six different micro-UAV controllers. As it is clear from the figure, each micro-UAV signal has a different waveform which can be attributed to the unique characteristics of the transmitter circuits, modulation techniques, and the packet structure. This makes it unreliable to use simple thresholding techniques to detect micro-UAVs especially in noisy environments. It is also difficult to detect the time-domain transient to obtain the fingerprint of the corresponding control signal because the end point of the transient is not clear unless there is an overshoot in the signal. 

In order to perform detection, all non-UAV signals are classified as noise. This includes background noise in the receiver itself and interference from all other wireless sources such as Wi-Fi, Bluetooth, and microwave oven which operate in same frequency band. Thus, the overall goal of the detection algorithm is to classify the received signals as belonging to the UAV or non-UAV class. However, due to the lack of data for the non-UAV class at this point of a time, we restrict our discussion to the classification of UAV and noise signals, albeit, the proposed method is developed for a more general case. Other wireless sources will be explored in our future work. Therefore, the detection problem boils down to detecting the presence/absence of a UAV based on the received RF signals in a noisy environment.

A pre-processing process is applied to the RF signals before proceeding with the detection algorithm. The RF signals are transformed into the wavelet domain using a pre-defined wavelet tree. The rationale behind the adapted technique \cite{Aly06elsherbeni} is that it helps in detecting the possible RF signals even in the low SNR regime. This leads to a better detection ability which is a necessity for applications like micro-UAV threat detection.

\subsection{Pre-processing Step}
The SNR of the received RF signal varies with the distance and transmitted power of the wireless source. Higher SNRs can be achieved in multiple ways \cite{Ciancio:2006,Aly06elsherbeni}. One such way is the use of the wavelet transform for removing the bias in the received RF signals and de-noising up to a certain extent. \cite{Aly06elsherbeni,6195803,Hassan:2010:AMR:1840667.1928457}. The use of wavelets provides the two-fold advantage compared to the traditional time domain and Fourier domain analysis \cite{strang1993wavelet}. One is the improvement in the SNR (de-noising) and other in the data compression without loss of information. The former aspect is important to improve the detection capability. The latter aspect is important for low complexity system design which result in faster detection algorithms.

\begin{figure}[t!]
    \centering
    \includegraphics[scale=0.25]{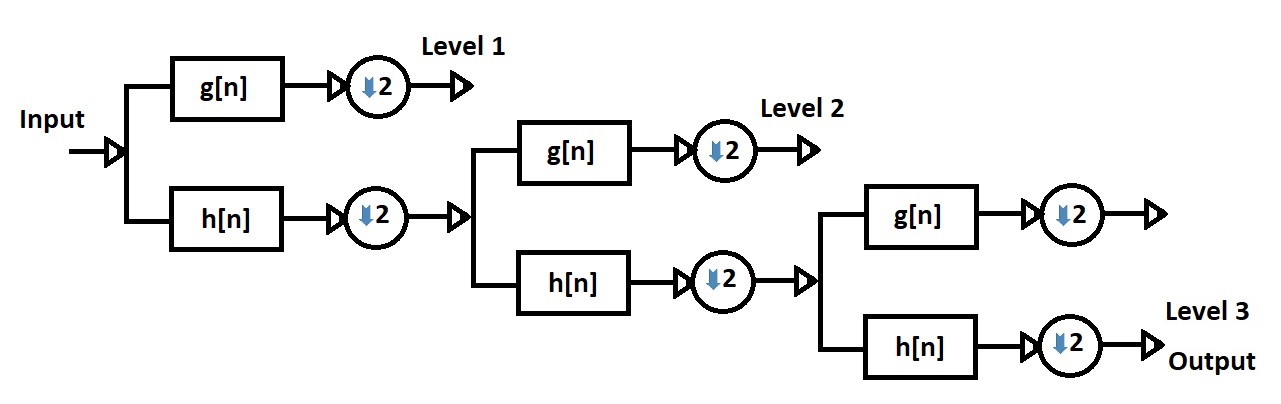}
    \caption{Three-stage wavelet decomposition.}
    \label{fig:Wavelet}
\end{figure}

\begin{figure}[b!]
    \centering
    \includegraphics[scale=0.35]{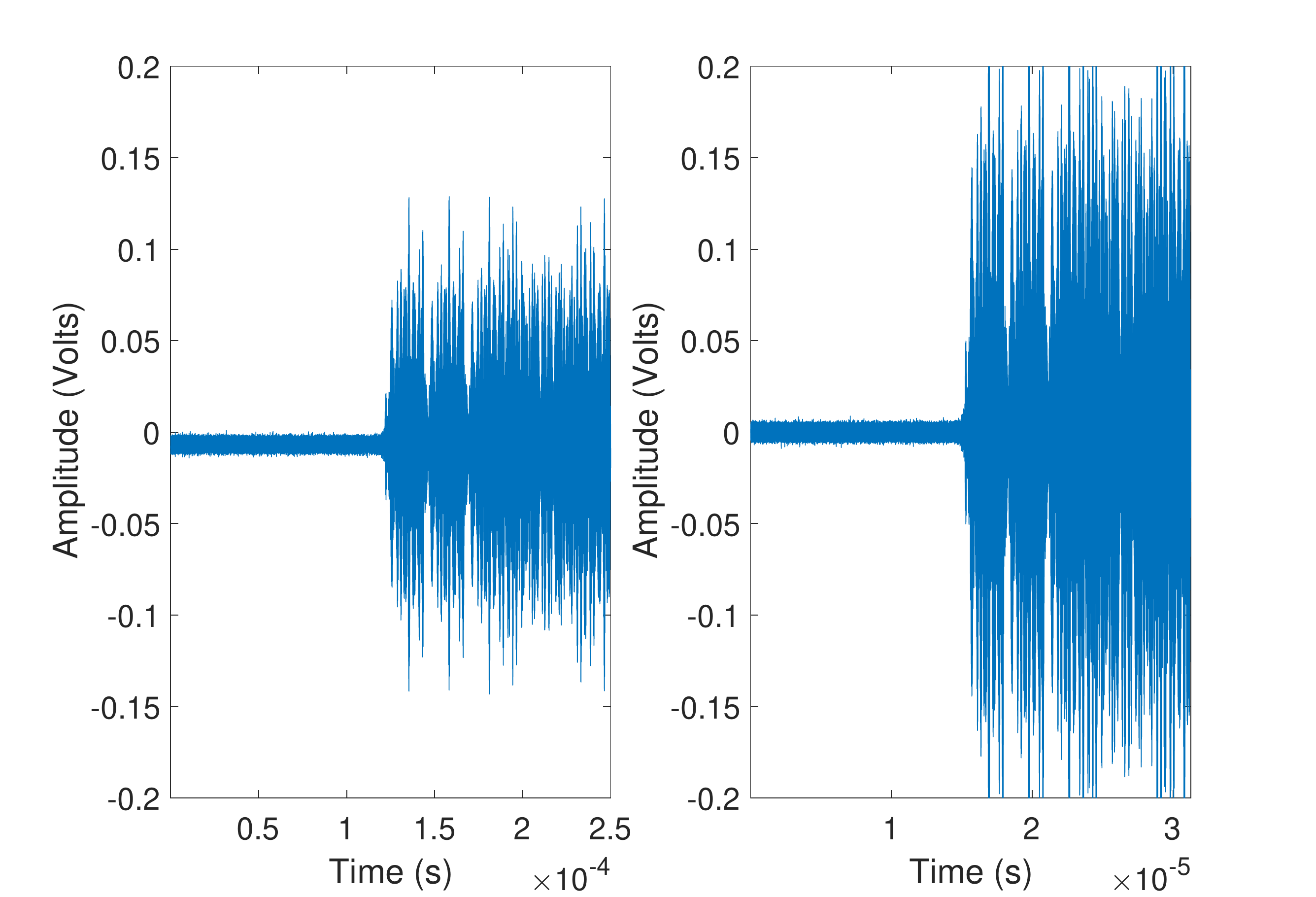}
    \caption{The raw signal from DJI Matrice 100 (left) and the corresponding output (right) obtained at the third stage of the wavelet decomposition.}
    \label{fig:Haart_Example}
\end{figure}

In this study, a three-stage wavelet decomposition is used as shown in Fig. \ref{fig:Wavelet}. Low-pass $g[n]$ and high-pass $h[n]$ filters of the Haar transform are chosen due to their simplicity. Each filter is followed by a down sampler. 
Wavelet coefficients obtained after the third level are considered as inputs to the detection algorithm both for training and testing the RF signals. An example of the RF signal received from the micro-UAV controller of DJI Matrice 100 and the corresponding wavelet transformed signal is shown in Fig. \ref{fig:Haart_Example}. It is clear from the figure that the wavelet transform removes the bias and reduces the number of samples while preserving the structure associated with the original raw signal.

\subsection{Bayesian Decision Making}
We use a Bayesian approach for the decision making process. To state the general problem, let ${C} \in \{0,1\}$ be an index denoting the class of the measured RF signal $y$. When $y \in$ UAV class then ${C} = 1$; otherwise ${C} = 0$. Let $y_{T}$ be a vector containing the transformed RF signal $y$. 
Then, the posterior probability of the UAV class given $y_T$ is
\begin{equation}
    P({C}=1|y_{T}) = \frac{P(y_T | {C}=1) P({C}=1)}{P(y_T)},
\end{equation}
where $P(y_T | {C}=1)$ is the likelihood function conditioned on ${C}=1$, $P({C} = 1)$ is the prior probability of UAV class, and $P(y_T)$ is the evidence. A similar expression holds for $P({C}=0|y_T)$. 
In terms of posterior probabilities, we decide ${C} = 1$, if
\begin{equation}
    P({C}=1|y_{T}) \geq P({C}=0|y_{T}).
\end{equation}
If we assume the number of samples from each of the classes in the training set are equal, the prior probabilities of the UAV and noise class become equal. Thus, the decision is favored to ${C}=1$, if
\begin{equation}
    P(y_{T}|{C}=1) \geq P(y_{T}|{C}=0).
\end{equation}
While this simplifies the problem of making a decision, there is still the problem of computing the likelihood $P(y_T|{C}=\{0,1\})$. This calculation is central to any Bayes decision problem because it reflects the interdependence of the classes of nature. In order to capture the dependency between the states, we incorporate the method discussed in the next subsection. We will get back to the calculation of the likelihood after we introduce the concept of states.

A close inspection of the collected data revealed that most of the UAV signals are differently structured (see Fig. \ref{fig:UAV data waveforms}). This is true for signals from other wireless sources as well. However, the same cannot be said about the noise data. That is, often the UAV data changes smoothly resulting in consecutive states of the signal that are not statistically independent. 
This is particularly useful when the SNR is low because at low SNRs the measured signal looks like a random noise.

\begin{figure}[b!]
    \centering
    \includegraphics[scale=0.2]{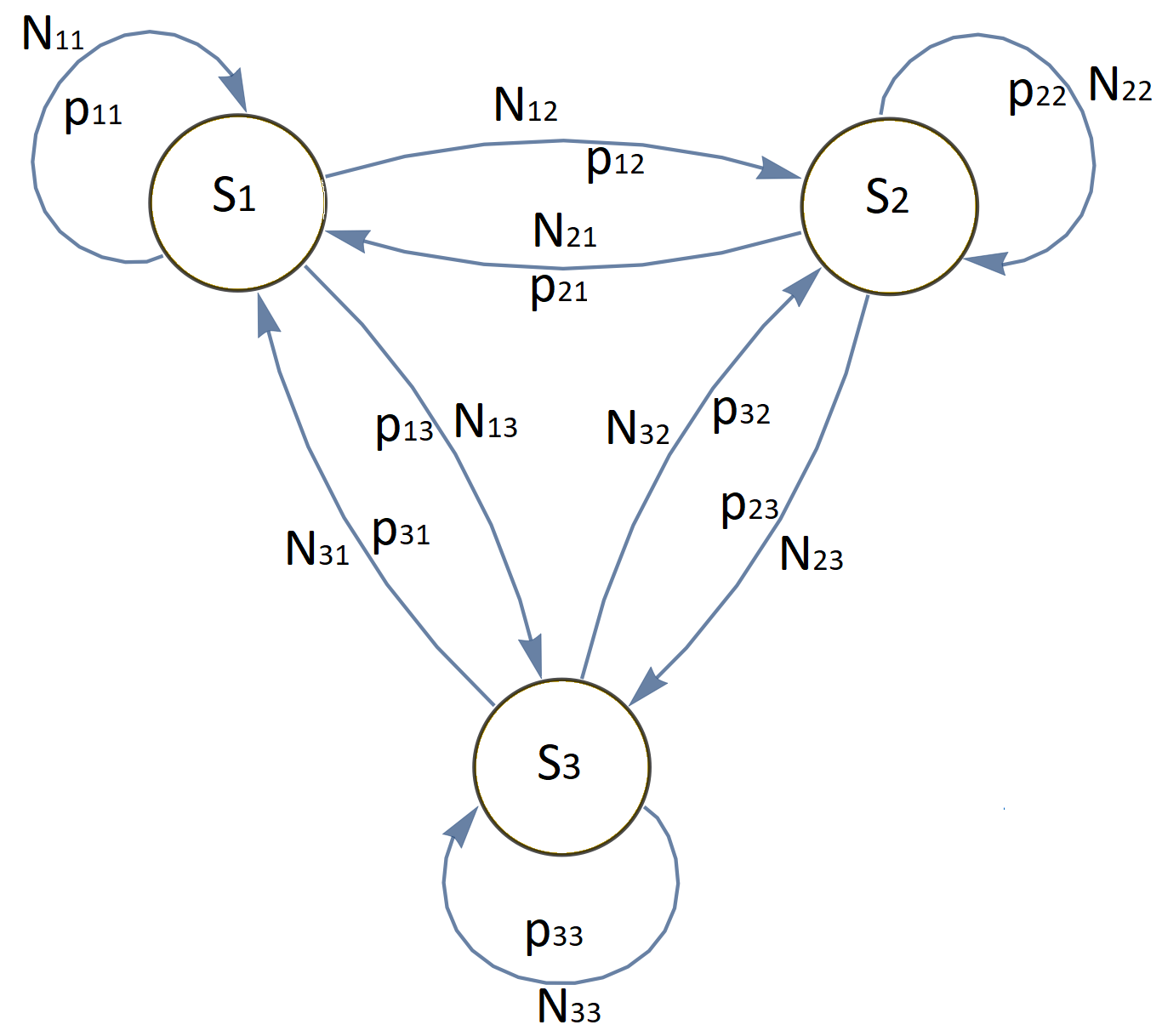}
    \caption{Definition of transition numbers and probabilities between the states.}
    \label{fig:transition}
\end{figure}

In order to exploit the dependency between the adjacent states, we define 3-state Markov models for each class.
We define two thresholds $T_1$ and $T_2$ based on the amplitude of the wavelet transformed noise signal, which are used to distinguish the three states. The values of $T_1$ and $T_2$ (with $T_1 > T_2$) are fixed based on the training data. We present the rationale of the choice of $T_1$ and $T_2$ in the subsequent subsections and validate the choice in the numerical simulation section.

With the help of the above thresholds, we define three-states $S_1$, $S_2$, and $S_3$. Based on the amplitude of the signal at each index $n$, a state is allotted to that index based on the following decision rule:
\begin{equation}
{S}_{y_T}(n)=
\begin{cases}
S_1, &  y_T[n]>T_1 \\
S_2, & T_2 \leq y_T[n] \leq T_1 \\
S_3, & y_T[n]< T_2 
\end{cases},
\end{equation}
where ${S}_{y_T}(n)$ is a set containing the states of the signal. Based on the above rule, it is straightforward to obtain the states associated with any time series signal. 
Once the state sequence ${S}_{y_T}$ is obtained, the probability of a transition between any two states is calculated. The transition probability matrix are generated based on the transitions from adjacent indexed states as seen in Fig. \ref{fig:transition}. The transition number and probability matrix are defined as follows:
\[
T_N = 
\begin{bmatrix}
    N_{11}       & N_{12} & N_{13}  \\
    N_{21}       & N_{22} & N_{23}  \\
    N_{31}       & N_{32} & N_{33} 
\end{bmatrix}, 
\]
\vspace{-0.45cm}
\[
T_P = 
\begin{bmatrix}
    p_{11}       & p_{12} & p_{13}  \\
    p_{21}       & p_{22} & p_{23}  \\
    p_{31}       & p_{32} & p_{33} 
\end{bmatrix} = \frac{T_N}{\sum_{i,j}^{} N_{ij}},
\]
respectively, where $T_\text{N}$ is a matrix capturing the number of transitions between any two states, $N_{ij}$ is the number of samples transiting from state $S_i$ to $S_j$; the matrix $T_P$ captures the transition probabilities and is obtained upon normalizing the $T_N$ matrix with the total number of samples. Here, $p_{ij}$ is the transition probability from state $S_i$ to $S_j$ for the signal of interest, i.e., $p_{ij} = p(S_i \rightarrow S_j)$. Note that the transition probabilities generated for the UAV and the noise data will be different when the considered SNR level is modest. Also, note that the choice of $T_1$ and $T_2$ dictates the transition probabilities for both the noise and UAV class data.

The choice of the thresholds has a direct impact on the decision making process. $T_1$ and $T_2$ are set to $\pm 3 \sigma$ of the wavelet transformed environmental noise signal, where $\sigma$ represents the standard deviation. The basis for this choice is that the environmental noise often is modeled as a Gaussian noise and the noise samples will be within the $\pm 3 \sigma$ band with a very high probability ($\approx 0.993$). The validation of the choice is presented in the numerical simulation section. Based on these settings, the UAV and noise class training transition probabilities are calculated.

The following procedure is followed to obtain the UAV transition probability matrix. The desired UAV packets from all the classes are appended and the states and transition probabilities are generated. A similar approach is employed to calculate the noise transition probabilities based on the collected environmental noise data.

For a given test signal, the signal goes through the similar pre-processing steps. Based on the pre-processed output, the states are defined and the $T_N$ is calculated. Finally, the likelihood of the class being a UAV is calculated as follows:
\begin{eqnarray}
\begin{aligned}
      P(y_T|{C}=1) &= \sum_{i,j=\{1,2,3\}:} P(S_i \rightarrow S_j | {C}=1),\\
           &= \prod_{i,j=\{1,2,3\}:} T_{P_{{C}=1}}(i,j)^{T_N(i,j)},\\
     &= \prod_{i,j=\{1,2,3\}:} p_{11;{{C}=1}}^{N_{11}}p_{12;{{C}=1}}^{N_{12}}\ldots p_{33;{{C}=1}}^{N_{33}}.
\end{aligned}
\end{eqnarray}
The log-likelihood of the above expression results in 
\begin{eqnarray}
\begin{aligned}
     \log\left( P(y_T|{C}=1)\right) &= \sum_{i,j=\{1,2,3\}:} N_{ij} \log(p_{ij;{{C}=1}}).
\end{aligned}
\end{eqnarray}
Similarly, the log-likelihood of the signal coming from a noise class is calculated by
\begin{eqnarray}
\begin{aligned}
     \log\left( P(y_T|{C}=0)\right) &= \sum_{i,j=\{1,2,3\}:} N_{ij} \log(p_{ij;{{C}=0}}).
\end{aligned}
\end{eqnarray}
The decision will be favored to $ {C}=1$, if $\log(P(y_T| {C}=1)) \geq \log(P(y_T|{C}=0))$; otherwise, ${C}=0$. We discuss the detection results in the simulation section for different SNR values. If the signal of interest is from the UAV class, then the classification stage is invoked. 

\section{UAV CLASSIFICATION USING RF FINGERPRINTS}\label{four_1}


We propose a technique based on the energy-time-frequency domain. Energy transients extracted in this domain can be used as the fingerprints of the corresponding signals.

\begin{figure}[t!]
\center
\includegraphics[scale=0.55]
{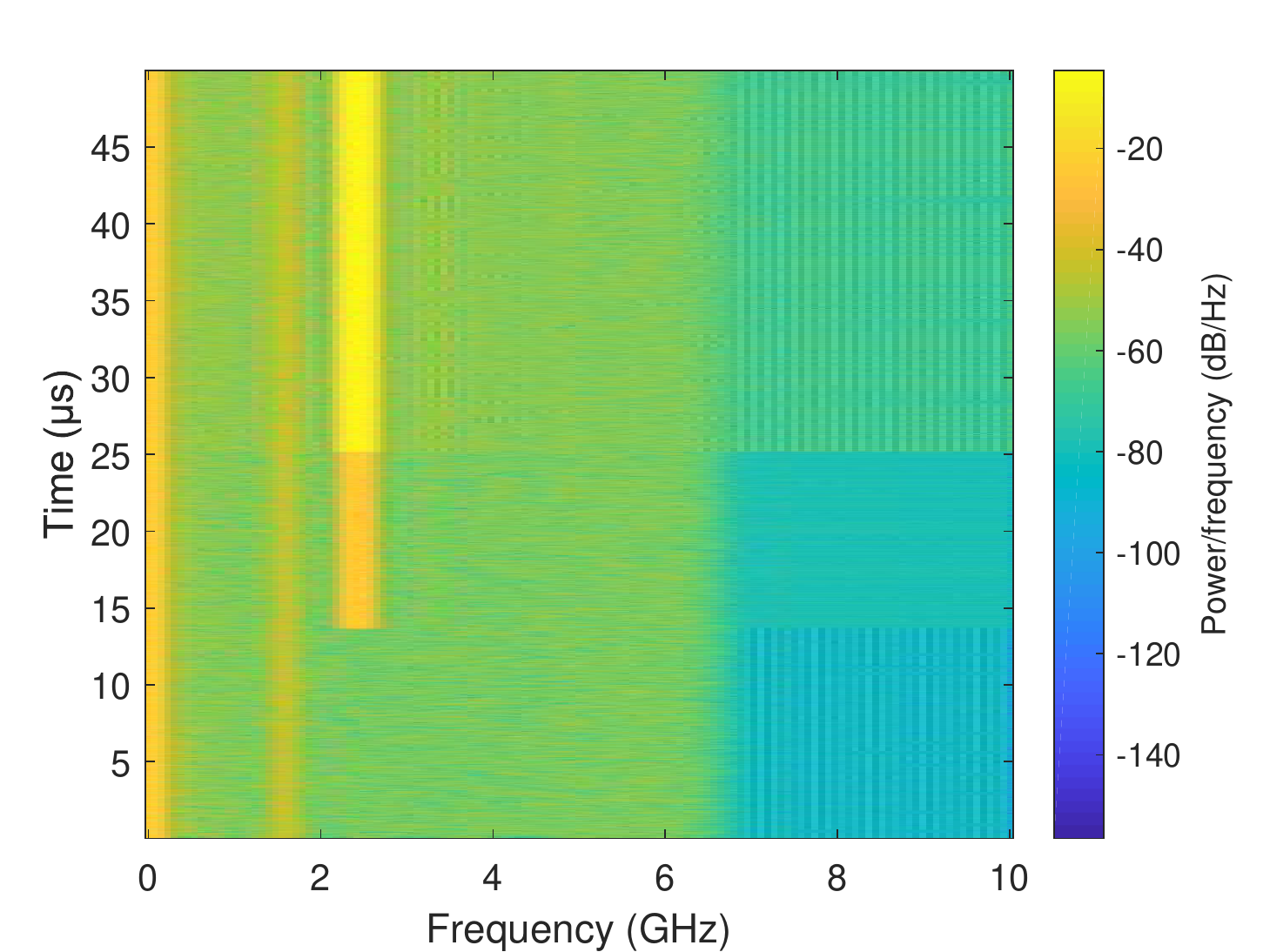}
\caption{Spectrogram of the RF signal shown in Fig.~\ref{fig:Haart_Example}.}
\label{Fig:spectrogram}
\end{figure}

For the representation of the RF signals in the energy-time-frequency domain, we use the spectrogram method. The spectrogram of any signal is computed using the squared magnitude of the discrete time short-time Fourier transform~(STFT)\\
\vspace{-0.25cm}
\begin{equation}
     \text{Spectrogram} (n,\omega) = \left| \sum_{n=-\infty}^{\infty}y[m]\text{w}[n-m]e^
     {-j\omega n} \right|^{2},
\end{equation}
where $y[n]$ is the captured signal, $m$ is discrete time, $\omega$ is the frequency, and w[n] is a sliding window function that acts as a filter. In addition, the spectrogram analysis of the captured RF signals can reveal the transmit frequency of the signal as well as the frequency hopping patterns. These are vital detection information. The spectrogram of the RF signal in Fig.~\ref{fig:Haart_Example} is shown in Fig.~\ref{Fig:spectrogram}. It can be clearly seen that the transmit frequency of the signal is 2.4 GHz.

\begin{figure}[h!]
 \center
 \includegraphics[scale=0.55]{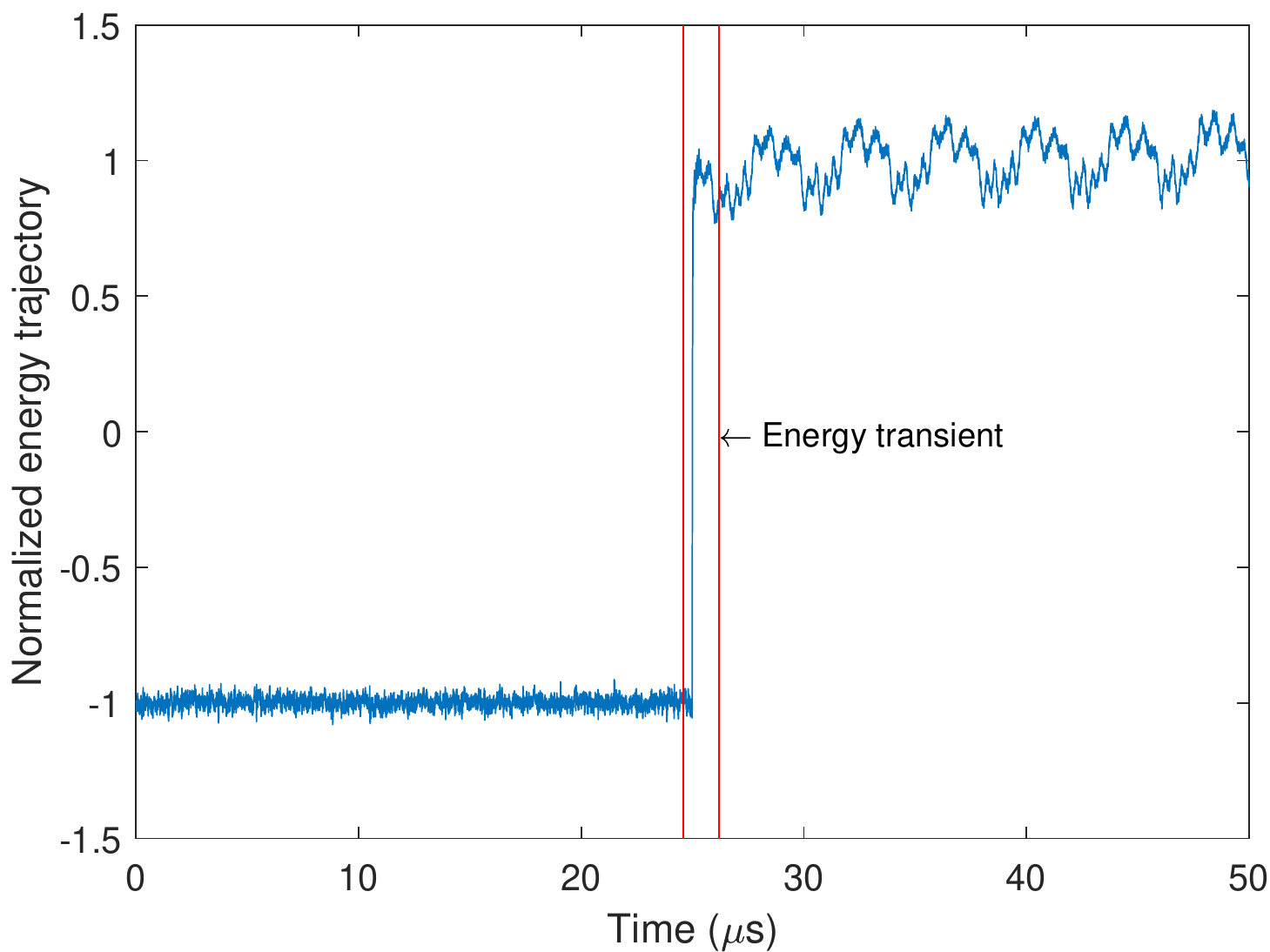}
\caption{Energy trajectory computed from the spectrogram in Fig.~\ref{Fig:spectrogram}. }
\label{Fig:Energy_Trajectory}
\end{figure}

The spectrogram, by definition, displays the energy/intensity distribution of the signal along the time-frequency axis. Therefore, the energy trajectory can be computed from the spectrogram by taking the maximum energy values along the time-axis. From this distribution, we estimate the energy transient by searching for the most abrupt change in the mean or variance of the normalized energy trajectory. The energy transient defines the transient characteristics of the signal in energy domain and is represented by $f_{E}(n)$, $n = 1,\dotsc ,N$. 
 For the RF signal in Fig.~\ref{fig:Haart_Example}, the normalized energy trajectory computed from the spectrogram, and the corresponding energy transient are shown in Fig.~\ref{Fig:Energy_Trajectory}.

\begin{figure}[b]
\center
\includegraphics[scale=0.55]{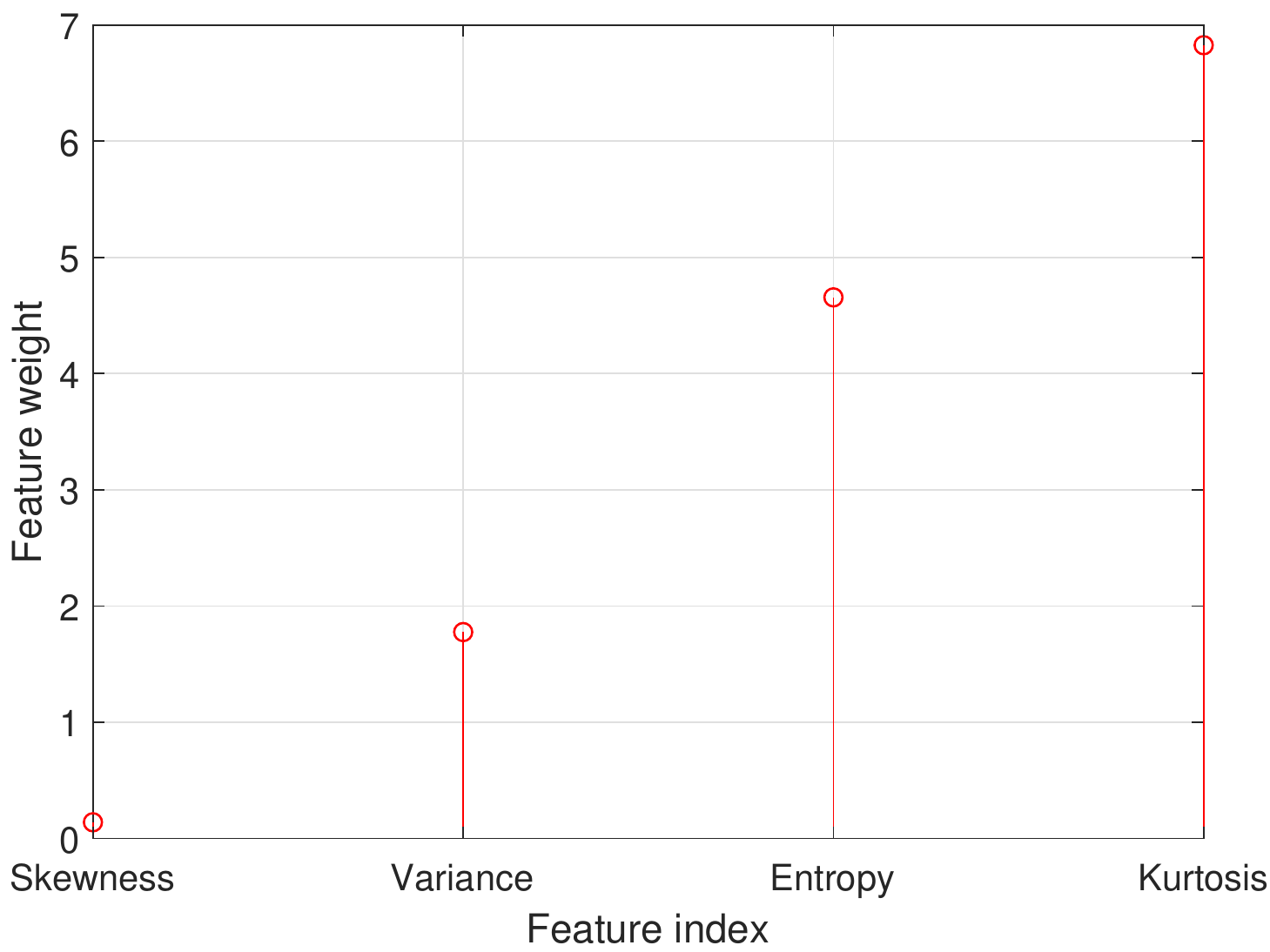}
\caption{NCA results showing the relative importance of the statistical features.}
\label{Fig:NCA_100samples}
\end{figure}
Once the energy transient is detected, the RF fingerprints are extracted. These fingerprints are the statistical moments that describe the energy transient. The extracted features are skewness ($\gamma$), variance ($\sigma^2 $), energy spectral entropy ($H$), and kurtosis ($k$). Physically, $\gamma$ is a measure of the asymmetry of the energy distribution around the mean value; $\sigma^2$ measures the spread of the energy trajectory about the mean value; $H$ provides a measure of the Shannon entropy (energy spectral power), and $k$ is a measure of the sharpness or flatness of the energy transient. These features are defined in terms of the mean ($\mu$) and standard deviation ($\sigma$) of $f_{E}$ as follows
\begin{eqnarray}
\begin{aligned}
    \label{eq:6}
\gamma({f_E}) &=\frac{1}{N \sigma^3} \sum_{n=1}^{N} \left(f_{E}(n)-\mu\right)^3\\
\sigma^2({f_E}) &=\frac{1}{N} \sum_{n=1}^{N} \left(f_{E}(n)-\mu\right)^2\\
H({f_E}) &= -\sum_{n=1}^{N} f_{E}(n)\log_{2}f_{E}(n)\\
k({f_E}) &= \frac{1}{N \sigma^4} \sum_{n=1}^{N} \left(f_{E}(n)-\mu\right)^4.
\end{aligned}
\end{eqnarray}
The feature sets consisting of the above statistical parameters are used to train four popular machine learning algorithms: kNN, discriminant analysis (DA), SVM, and neural networks (NN)~\cite{pattern_recognition_book}. Since some of the features may be correlated, and so redundant, we perform feature selection to reduce the computational cost of the classification algorithm. This is discussed next.

\subsection{Feature Selection Using NCA}
 In practice, it is often required to reduce the dimensionality of a feature set by removing correlated features. 
By this means, computational cost of the classification algorithm can be reduced. Most often, a feature selector is a linear operator that projects the original data or feature set into a lower dimensional space. Neighborhood component analysis (NCA) is such a linear projector. 
 
NCA is a non-parametric, embedded, and supervised learning method for feature selection. NCA learns a matrix by which the primary data are transformed into a lower-dimensional space~\cite{goldberger2005neighbourhood}. In this lower-dimensional space, the features are ranked according to a weight metric, with the more important features receiving higher weight values. NCA achieves this goal by maximizing the regularized objective function $f(\textbf{w})$ with respect to the weight variable \textbf{w}. The regularized objective function is defined as:
\begin{eqnarray}
\begin{aligned}
\label{eq:9}
     f(\textbf{w})&=\frac{1}{N}\sum_{i=1}^{N}{p_{i}}-\lambda\sum_{r=1}^{p}{{\bf{w}^2_r}},
\end{aligned}
\end{eqnarray}
where $\lambda$ is the regularization term, $N$ is the number of samples in the feature set, and $p_{i}$ is the average leave-one-out probability (LOO). In other words, $p_{i}$ is the probability that NCA correctly learns an observation in the feature set. In order to perform feature selection, NCA uses the regularization term to drive to zero all the weights corresponding to the redundant or correlated features. In~\cite{goldberger2005neighbourhood}, NCA is compared with the linear dimensionality reduction (LDA) on several dataset. It is observed that if the classes are not convex and cannot be linearly separated, then LDA result will be inappropriate. In contrast, NCA adaptively finds the best project matrix without assuming any parametric structure in the lower dimensional space. In the same experiment, NCA was shown to outperform relevant components analysis (RCA) and principal component analysis (PCA). 

NCA ranks the features according to their importance. Fig.~\ref{Fig:NCA_100samples} shows the results of the NCA performed for the data to be mentioned in Section~\ref{four}. Fig.~\ref{Fig:NCA_100samples} shows the result of NCA. The RF fingerprints are ranked according to their weight value. As can be seen, kurtosis has the highest weight and so is the most important RF fingerprint for this test case. On the other hand, skewness has the lowest weight and so is the least important RF fingerprint. This behavior is reasonable because there is a correlation between the features skewness and kurtosis. Consequently, for the training and testing, the classifiers can discard skewness and still produce good results. This can prevent the over-fitting problem when training the classifiers. In addition, for large-scale classification problems, there can be huge computational saving in training the classifiers with fewer number of significant features.

\begin{table*}[h!]
\centering
\caption{UAV Controller Catalogue.}
\label{Table:Table1}
\begin{tabular}{|P{1.5cm}|P{1.75cm}|P{1.75cm}|P{2cm}|P{1.75cm}|P{1.75cm}|P{1.75cm}|P{1.75cm}|}
\hline
\textbf{ID}& 1 & 2 & 3 & 4 & 5 & 6 & 7  \\
\hline
\textbf{Model Name} & DJI Inspire 1 & DJI Matrice 100 & DJI Phantom 4Pro  & DJI Phantom 3 & DX5e Spektrum & DX6e Spektrum & FlySKy FS-T6  \\  \hline
\end{tabular}\\
\vspace{0.25cm}
\begin{tabular}{|P{1.5cm}|P{1.75cm}|P{1.75cm}|P{2cm}|P{1.75cm}|P{1.75cm}|P{1.75cm}|P{1.75cm}|}
\hline
\textbf{ID}& 8 & 9 & 10 & 11 & 12 & 13 & 14 \\
\hline
\textbf{Model Name} &Futaba T8FG & Graupner MC-32  & Hobbie King-T6A V2 & JR X9303 & DX6i Spektrum & Turnigy 9X & Jeti Duplex DC-16 \\  \hline
\end{tabular}
\end{table*}

\begin{figure}[t!]
\center
\includegraphics[scale=0.6]
{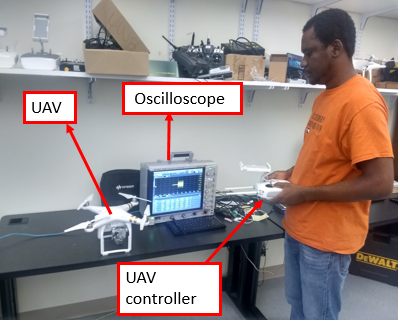}
\caption{Near-field measurement setup.}
\label{Fig:system_nearfield}
\end{figure}

\begin{figure}[t!]
\center
\includegraphics[scale=0.6]
{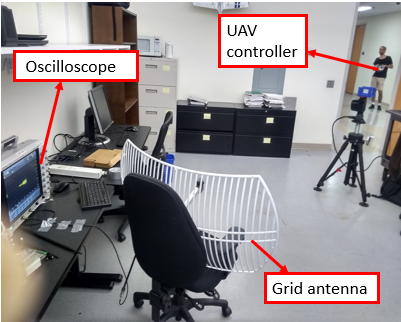}
\caption{Far-field measurement setup.}
\label{Fig:system_farfield}
\end{figure}


\begin{figure}[t!]
 \center
\includegraphics[scale=0.55]{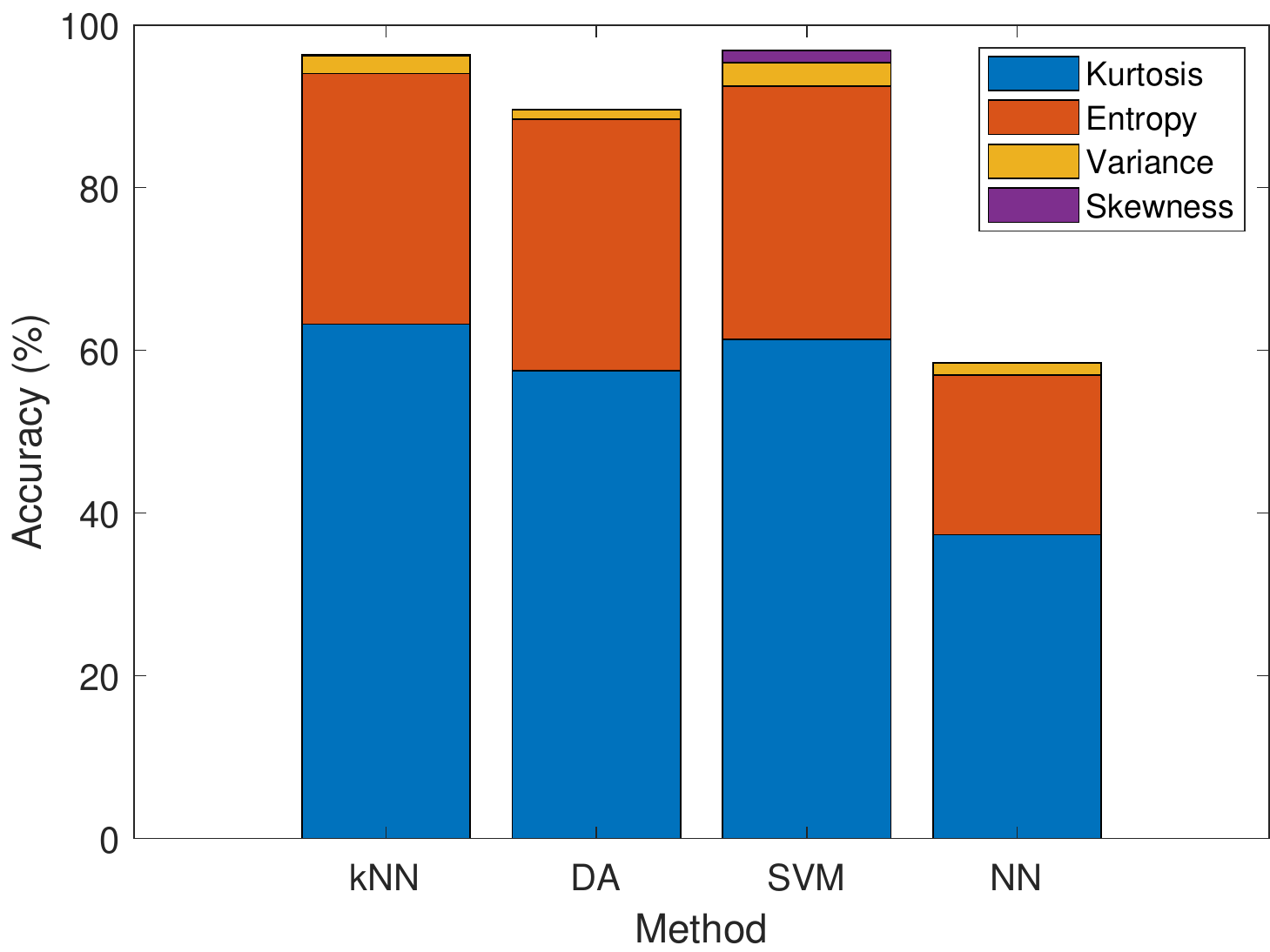}
 \caption{Impact of the features on the classification accuracy of the different machine learning methods.}\label{Fig:Accuracy_VS_different_features}
\end{figure}

\begin{figure}[t!]
 \center
 \includegraphics[scale=0.55]{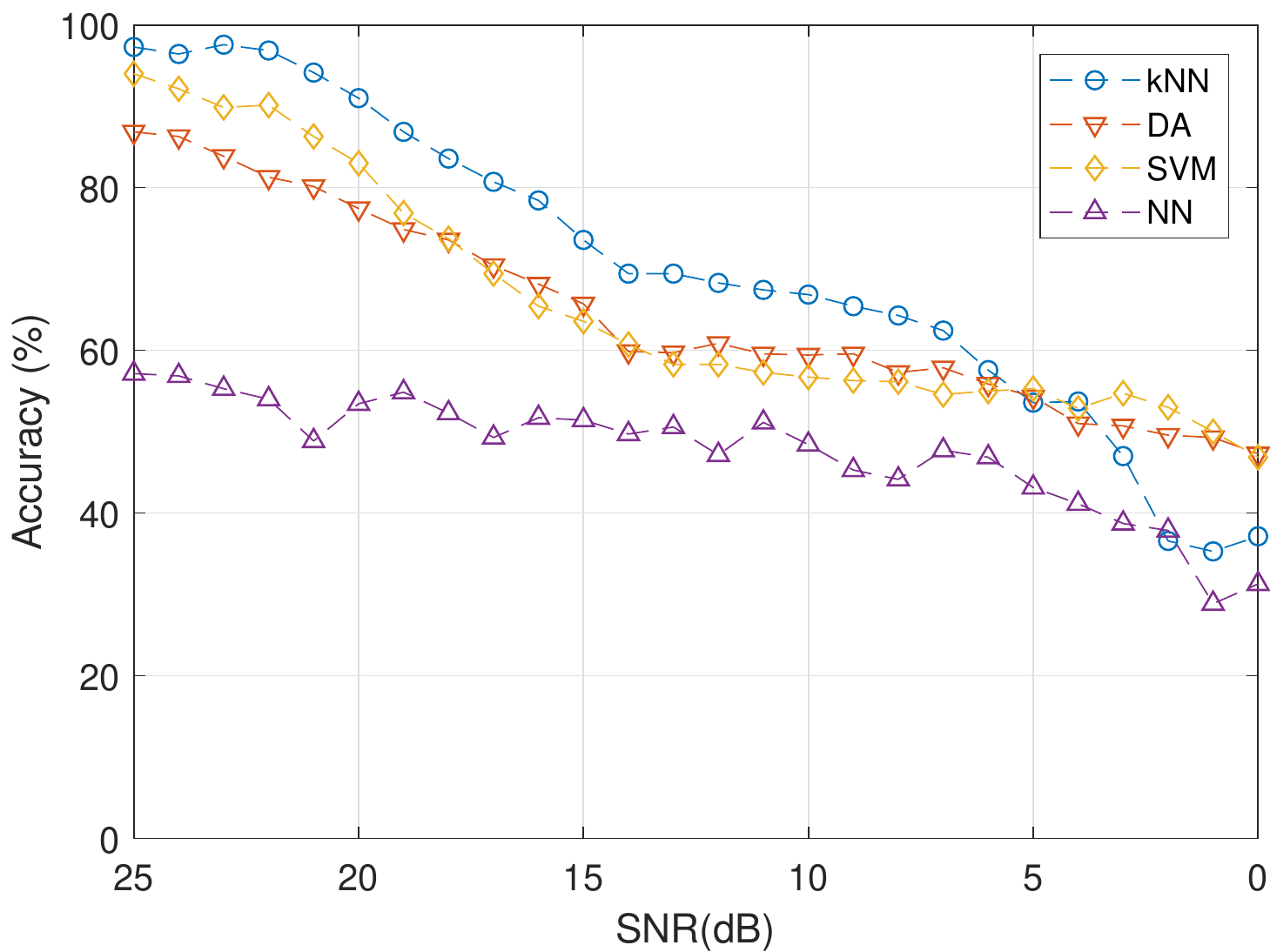}
 \caption{Classification accuracy versus SNR.}
 \label{Fig:Accuracy_VS_SNR}
\end{figure}

\begin{table*}[t!]
\centering
\caption{Detection Performance.}
\label{Table:Detection_performance}
\begin{tabular}{|P{1.75cm}|P{0.75cm}|P{0.75cm}|P{0.75cm}|P{0.75cm}|P{0.75cm}|P{0.75cm}|P{0.75cm}|P{0.75cm}|P{0.75cm}|P{0.75cm}|P{0.75cm}|P{0.75cm}|P{0.75cm}|P{0.75cm}|P{0.75cm}|P{0.75cm}|}
\hline
\multicolumn{1}{|c}{\multirow{1}{*}{}}&\multicolumn{13}{|c|}{\textbf{SNR (dB)}}\\
\hline
 & 0 & 2 & 4 & 6 & 8 & 10 & 12 & 14 & 16 & 18 & 20 & 22 & 24 \\
\hline
\textbf{Detection Accuracy ($\%$)} &13 &19  &23 &46  & 61  &84  & 100 &100  & 100  &100  & 100  &100  &100 \\  \hline
\end{tabular}
\end{table*}

\begin{figure}[t!]
 \center
 \includegraphics[scale=0.55]{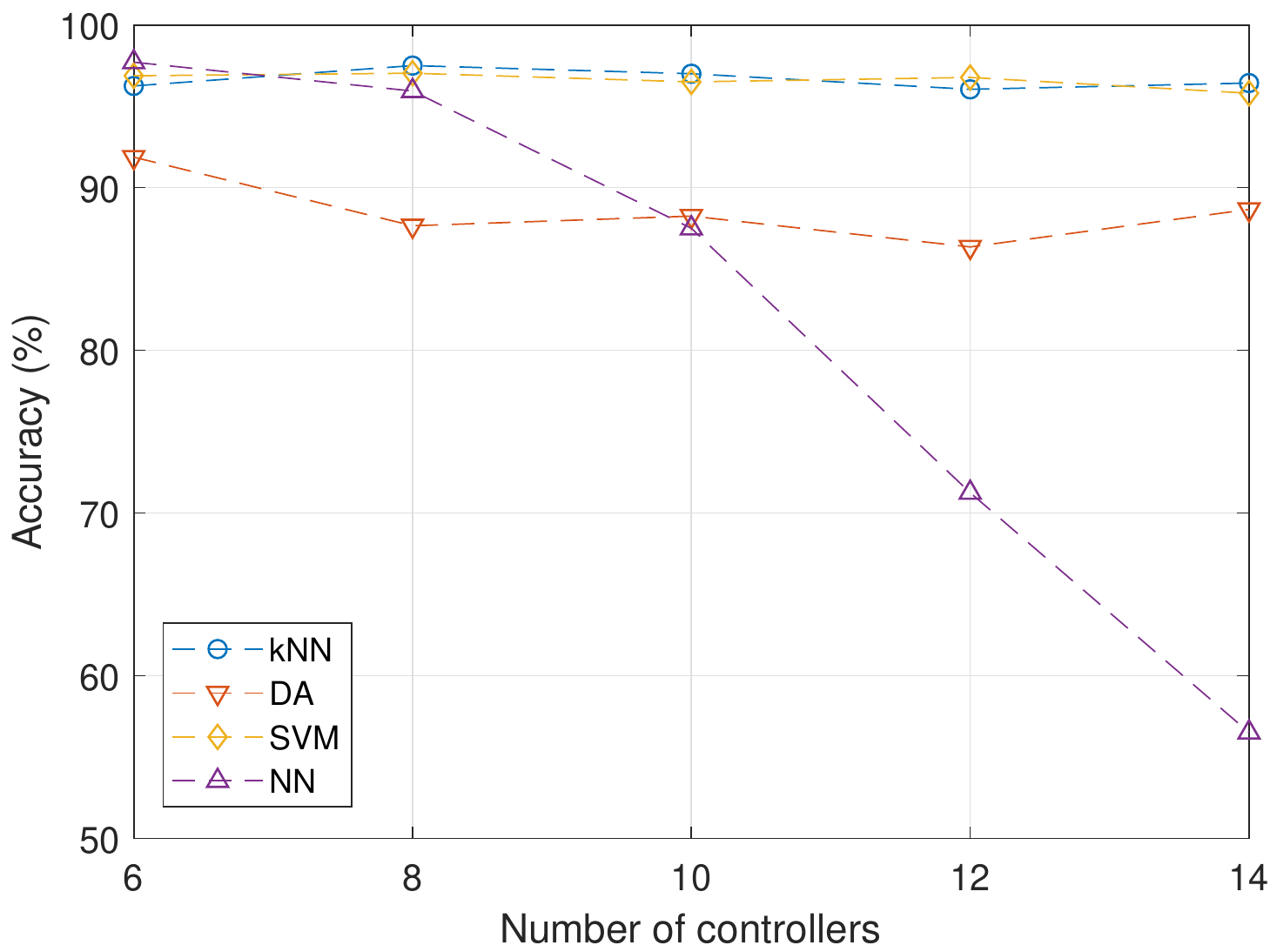}
 \caption{Classification accuracy versus the number of controllers at SNR of 25 dB.}
 \label{Fig:Accuracy_VS_numberOfControllers}
\end{figure}


\section{EXPERIMENTS AND RESULTS}\label{four}

\subsection{Experimental Setup} 
RF signals are collected indoor from 14 micro-UAV controllers operating at 2.4 GHz in near-field. Table~\ref{Table:Table1} gives the catalogue of the micro-UAV controllers used for data collection and their class label (ID). The indoor experimental environment is very noisy with strong interference from several sources operating in the 2.4 GHz frequency band: Wi-Fi, Bluetooth and micro-wave oven. The experimental setup for both near-field and far-field indoor scenarios are shown in Figs.~\ref{Fig:system_nearfield} and~\ref{Fig:system_farfield}, respectively. The RF sensing and detection system consists of a 6 GHz bandwidth Keysight MSOS604A oscilloscope with the highest sampling frequency of 20 Gsa/s, 2 dBi omnidirectional antenna, and 24 dBi Wi-Fi grid antenna. The antennas operate in the 2.4 GHz frequency band. The omnidirectional antenna is used to capture the UAV controller RF signal at close distance while the grid antenna is used for far-field signal capture (reduced SNR scenarios). At the near-field, the SNR is about 30 dBi and decreasing with the distance from the receiver.

\begin{table*}[h!]
\centering
\caption{Confusion Matrix for the kNN Method Computed From 280 Test Signals Obtained From 14 Different Micro-UAV Controllers at SNR of 25 dB.}
\begin{tabular}{|P{0.1cm}|P{0.6cm}|P{0.6cm}|P{0.6cm}|P{0.6cm}|P{0.6cm}|P{0.57cm}|P{0.57cm}|P{0.57cm}|P{0.57cm}|P{0.57cm}|P{0.57cm}|P{0.57cm}|P{0.57cm}|P{0.57cm}|P{0.57cm}|c|}
\hline
\multicolumn{1}{|c|}{\multirow{1}{*}{}}&\multicolumn{15}{|c|}{\textbf{Target Class}}&\multicolumn{1}{c|}{}\\ \hline
\multicolumn{1}{|c|}{\multirow{16}{*}{\rotatebox{90}{\textbf{Predicted Class}}}} &\textbf{ID}&\textbf{1}&\textbf{2}&\textbf{3}&\textbf{4}&\textbf{5}&\textbf{6}&\textbf{7}&\textbf{8}&\textbf{9}&\textbf{10}&\textbf{11}&\textbf{12}&\textbf{13}&\textbf{14}&\textbf{Out} ($\%$)\\ \cline{2-17} 
&\textbf{1}& 20 & 0 & 1  & 0  & 0 & 0 & 0 & 0 & 0  & 0 & 0 & 0 & 0 & 0 &95.2\\ \cline{2-17} 
&\textbf{2}& 0 & 20 & 0  & 0& 0 & 0 & 0 & 0 & 0  & 0& 0 & 0 & 0   & 0 & 100 \\ \cline{2-17} 
&\textbf{3}& 0 & 0 & 19  & 0& 0 & 0 & 0 & 0 & 0  & 0& 0 & 0 & 0   & 0 & 100  \\ \cline{2-17} 
&\textbf{4}& 0 & 0 & 0  & 20    & 0   & 0 & 0 & 0 & 0  & 0 & 0 & 0 & 0   & 0 & 100 \\ \cline{2-17}
&\textbf{5}& 0 & 0 & 0 & 0 & 20   & 2 & 0 & 1 & 0  & 0  & 0  & 2 & 0   & 0 & 80.0  \\ \cline{2-17} 
&\textbf{6}& 0 & 0 & 0  & 0    & 0   & 18 & 0 & 0 & 0  & 0  & 0   & 0 & 0 & 0 & 100 \\ \cline{2-17} 
&\textbf{7}& 0 & 0 & 0  & 0    & 0   & 0 & 20 & 0& 0  & 0    & 0   & 0 & 0   & 0 & 100 \\ \cline{2-17} 
&\textbf{8}& 0 & 0 & 0  & 0    & 0   & 0 & 0 & 18 & 0  & 0 & 0 & 0 & 0   & 1 & 94.7 \\ \cline{2-17} 
&\textbf{9} & 0 & 0 & 0 & 0 & 0 &0 &0 & 0 &20  & 0 & 0 & 0 & 0 & 0 & 100 \\ \cline{2-17} 
&\textbf{10}  & 0 & 0 & 0 & 0 &0  & 0 & 0 & 0 & 0  & 20 & 0 & 0 & 0   & 0 &100 \\ \cline{2-17} 
&\textbf{11}& 0 & 0 & 0  & 0 & 0  & 0 & 0 & 0 & 0  & 0 & 20 & 0 & 0   & 0 &100 \\ \cline{2-17} 
&\textbf{12}& 0 & 0 & 0  & 0 & 0  & 0 & 0 & 1 & 0  & 0 & 0 & 18 & 0   & 0 &94.7 \\ \cline{2-17} 
&\textbf{13}& 0 & 0 & 0  & 0 & 0  & 0 & 0 & 0 & 0  & 0 & 0 & 0 & 20 & 0 &100 \\ \cline{2-17} 
&\textbf{14}  & 0 & 0& 0 & 0 &0 & 0 & 0 & 0 & 0  & 0 & 0 & 0 & 0 & 19 &100  \\ \cline{2-17}
& \textbf{Out} ($\%$) & 100 & 100 & 95  & 100  & 100  & 90 & 100  & 90 & 100 & 100&100 & 90 & 100  & 95 &97.1  \\ \cline{1-17} 
\end{tabular}
\label{Table:confusion_Matrix_KNN1}
\end{table*}

The RF signals from the micro-UAV controllers are captured by the antenna and fed into the oscilloscope (receiver system). The collected data are automatically saved in a cloud database for post processing. For each controller, 100 RF signals are collected. Each RF signal is a vector of size $5000000 \times 1$ and has a time span of 0.25 ms. The database are partitioned with the ratio $p=0.2$. That is, 80\% of the saved data is randomly selected for training and the remaining 20\% is used for testing (4:1 partitioning). 

\subsection{Results}

During the experiments, the environmental noise was fairly static. However, there were significant scattering and absorption from objects in the hallway. We took measurements in the hallway at various distances up to 130 m. We observed the signal vanishes beyond 130 m. Thus, the hallway behaves like a lossy rectangular waveguide. In addition, it was observed that the polarization planes of the transmitter and receiver antennas significantly affected the received signal strength. It was noticed that $\pm 3\sigma$ of the Wavelet transformed environmental noise was around 0.0098 volts. Thus, throughout the work, $T_1$ and $T_2$ was set to $\pm 0.0098$ volts, respectively.

 The performance of the detection algorithm is presented in Table.~\ref{Table:Detection_performance}. As expected, we see that the detection accuracy increases with the SNR. When SNR=10 dB, which corresponds to a distance of 80 m, the detection accuracy is 84\%. The detection accuracy increases as we reduce the distance between the UAV controller and receiver antenna. The system is able to detect all the UAVs at any SNR values of beyond 12 dB.  
 The performance could be further improved by attaching an external low-noise power amplifier to the input of the oscilloscope. This will reduce the input noise from the environment and improve the detection accuracy.
  
 Once a UAV controller is detected, the received RF signal should be classified to identify which UAV it is. In order to validate the classification methods, 10 Monte Carlo simulations are run. The average accuracy of each method is calculated for a number of cases.  
Fig.~\ref{Fig:Accuracy_VS_different_features} shows the classification accuracy of each method as well as the impact of different feature selections on the performance of that method. kNN and SVM perform similarly with a classification accuracy of 96.3\% and 96.84\%, respectively, and are followed by DA with an accuracy of 88.15\%. NN can only achieve a classification accuracy of 58.49\% when there are 14 micro-UAV controllers. Fig.~\ref{Fig:Accuracy_VS_different_features} also verifies the results of NCA given in Fig.~\ref{Fig:NCA_100samples} where we see the relative importance of the features in the classification accuracy. From the figure, it is obvious that kurtosis is the most significant feature, that is, the feature contributing the most to the classification accuracy. As predicted by NCA, considering skewness in addition to the other three features contributes the least to the classification accuracy. This observation holds for all the methods.

Fig.~\ref{Fig:Accuracy_VS_SNR} shows the classification performance of the methods at different SNRs. As expected, the classification accuracy decreases as the distance between the UAV controller and the receiver system increases. At an SNR of 25 dB, corresponding to a distance of about 5 m, kNN achieves a classification accuracy of about 97.29\%. At this distance, DA and SVM show similar performance. However, NN achieves a classification accuracy of only 57.14\%. When the SNR is 10 dB, corresponding to a distance of about 80 m, kNN, DA and SVM achieves a classification accuracy between 60-70\% while NN performs below 50\%. 
  
 In Fig.~\ref{Fig:Accuracy_VS_numberOfControllers}, we investigate the robustness and stability of the classifiers as the number of UAV controllers increases. At SNR of 25 dB, we see that the performance of the kNN, DA, and SVM remains almost the same with the change in the number of the controllers. 
 However, NN shows instability when there are 8 or more controllers. It is clear that the NN is not a good choice unless there are 6 or less controllers at least with the feature set used in this study.

 It is obvious from the observations so far that kNN performs the best and NN performs the worst. This is probably because we did not optimize the hyper parameters of the NN algorithm. In general, NN algorithms are very sensitive to the choice of hyper parameters used. On the other hand, hyper-parameter optimization was built into the kNN, DA and SVM classifiers used in this study. Hyper-parameter optimization for the NN method will be investigated in our future works. Therefore, because of the superior performance of kNN in this study, it will be considered as the base classification method.
   
Table~\ref{Table:confusion_Matrix_KNN1} shows a sample confusion matrix obtained for the kNN classifier. This table describes the performance of the kNN model on a set of test data (RF signals) for which the true labels are known. The test data were captured at an SNR of 25 dB. The confusion matrix shows that kNN achieves an accuracy of 97.1\%. Except one or two samples from 4 controllers, the classsifier is not confused between the micro-UAVs.

\section{CONCLUSION}
\label{five}

In this paper, we investigated the problem of detecting and classifying micro-UAV control signals. The detection algorithm uses a Bayesian approach based on the Markov models of UAV and non-UAV classes while the classification method relies on the energy-time domain RF signal and uses features extracted in this domain. We show that the kNN classifier performs the best while NN has the the worst performance when considering lower SNR levels and increased number of controllers. We obtain an accuracy of above 80\% with the kNN classifier up to SNR of 15 dB for 14 controllers. We also show that it is possible to increase the number of controllers up to a certain level without compromising the performance using kNN and SVM methods where both result in accuracy of above 95\%. In the future work, we will perform experiments in outdoor environment using multiple sensors and UAV signals for micro-UAV detection and classification. 
This approach will be more effective in modern electronic warfare environment, where autonomous military UAVs employ low probability of intercept (LPI) emitters which are difficult to detect by a single RF sensing platform due to their low peak power. This problem can be addressed in our future work by using netted sensor fusion system and deep learning algorithms for cluster fingerprinting based detection and classification of these LPI emitters. 
Moreover, in such an advanced system, techniques for specific emitter identification (SEI) such as the formation of 3D feature cluster map could be investigated for improved classification. These are beyond the scope of the current work. 

\section*{ACKNOWLEDGMENT}
This work has been supported by the NASA grant NNX17AJ94A.

\bibliography{IEEEabrv,references}
\bibliographystyle{IEEEtran}

\end{document}